\documentclass[11pt]{article}
\usepackage[dvips]{graphics}
\usepackage{epsfig}
\usepackage{chicago}

\newcommand{\vc} {{\bf c}}
\newcommand{\vz} {{\bf z}}

\newcommand{\vx}  {{\bf x}}

\newcommand{\vq} {{\bf q}}

\begin{document}
\title{A Moving Bump in a Continuous Manifold: \\
A Comprehensive Study of
the Tracking Dynamics of Continuous Attractor Neural Networks}
\author{C. C. Alan Fung$^1$, K. Y. Michael Wong$^1$ and Si Wu$^{2,3}$
\\
$^1$Department of Physics, Hong Kong University
of Science and Technology,
\\Clear Water Bay, Hong Kong, China
\\$^2$Department of Informatics, University of Sussex,
Brighton, United Kingdom
\\$^3$Lab of Neural Information Processing, Institute of Neuroscience,
Shanghai, China}

\maketitle

\begin{abstract}
Understanding how the dynamics of a neural network
is shaped by the network structure,
and consequently how the network structure 
facilitates the functions implemented by the neural system,
is at the core of using mathematical models
to elucidate brain functions.
This study investigates the tracking dynamics
of continuous attractor neural networks (CANNs).
Due to the translational invariance of neuronal recurrent interactions,
CANNs can hold a continuous family of stationary states.
They form a continuous manifold
in which the neural system is neutrally stable.
We systematically explore how this property
facilitates the tracking performance of a CANN,
which is believed to have clear correspondence with brain functions.
By using the wave functions of the quantum harmonic oscillator as the basis,
we demonstrate how
the dynamics of a CANN is decomposed into different motion modes,
corresponding to distortions in 
the amplitude, position, width or skewness of the network state.
We then develop a perturbative approach
that utilizes the dominating movement of the network's stationary states
in the state space. This method allows us to approximate the network dynamics
up to an arbitrary accuracy depending on the order of perturbation used.
We quantify the distortions of a Gaussian bump during tracking,
and study their effects on the tracking performance.
Results are obtained 
on the maximum speed for a moving stimulus to be trackable 
and the reaction time for the network to catch up 
with an abrupt change in the stimulus.
\end{abstract}

\section{Introduction}

The brain performs computation by updating its internal states
according to dynamical rules
that are determined by both the nature of external inputs
and the structure of neural networks.
A neural system acquires its network structure
through either learning from experience, 
or inheriting from evolution or both.
Its impact on network dynamics is twofold:
on one hand, it determines the stationary states of the network
that lead to associative memory;
and on the other hand, it carves the landscape of the state space
of the network as a whole, 
which may contribute to other cognitive functions,
such as movement control, spatial navigation,
population decoding and object categorization.
Understanding how the dynamics of a neural network
is shaped by its network structure
is at the core of using mathematical models
to elucidate brain functions~\cite{Dayan01}.

Associative memories, such as those based 
on the Hopfield model~\cite{Hopfield84,Domany95},
have been intensively studied over the past decades.
However, an equally important issue
has not been investigated to the same extent,
namely, how a large-scale structure of the state space of a neural system
may facilitate brain functions.
Recently, a type of recurrent network,
called continuous attractor neural networks (CANNs),
has received wide attention~\cite{Wilson72,Amari77,Georgopoulos82,Maunsell83,Funahashi89,Wilson93,Rolls95,Ben-Yishai95,Zhang96,Seung96,Samsonovich97,Camperi98,Ermentrout98,Hansel1998,Taube98,Deneve99,Wang01,Laing01,Wu02,Stringer02,Brody03,Erlhagen03,Gutkin03,Renart03,Trappenberg03,Folias04,Wu05,Chow06,Miller06,Machens08,Wu08}.
These models are particularly useful when modeling
the encoding of continuous stimuli in neural systems.
Different from other attractor models,
CANNs have a distinctive feature
that their neuronal interactions are translationally invariant.
As a result, they can hold a family of stationary states
that can be translated into each other
without the need to overcome any barriers.
In the continuum limit, these stationary states
form a continuous manifold\footnote{A manifold is a mathematical concept.
Simply stated, it is a set of points with a coordinate system.
Here, each stationary state of the network is a point in the manifold.
As will be shown later,
these stationary states have the shape of a Gaussian bump,
and their peak positions define the coordinates.}
in which the system is neutrally stable,
and the network state can translate easily
when the external stimulus changes continuously.
Beyond pure memory retrieval,
this large-scale structure of the state space
endows the neural system with a tracking capability.
An illustration of the landscape of the state space of a CANN and how
it leads to neutral stability of the network stationary states
is shown in Fig.~1.

The tracking dynamics of a CANN has been theoretically studied
by several authors in the literature
(see, e.g.,\cite{Ben-Yishai95,Zhang96,Samsonovich97,Xie02,Folias04,Wu05}).
These studies have demonstrated clearly
that a CANN has the capacity of tracking a moving stimulus continuously
and that this tracking property can describe many brain functions well.
Detailed rigorous analyses of the tracking behaviors of a CANN
are still lacking, however, 
including, for instance,
1) the conditions under which a CANN can track a moving stimulus successfully;
2) how the network state is distorted during the tracking; and
3) how these distortions affect the tracking performance of a CANN.
In this study, we systematically investigate these issues.

We use a simple, theoretically solvable model of a CANN
as the working example.
The stationary states of the network have a Gaussian shape,
and the network dynamics can be described
using the basis functions of quantum harmonic oscillators.
These analytical results allow us
to explore the tracking performances of a CANN in much more detail
than in previous studies based on simulation observations.
In particular, we demonstrate clearly
how the dynamics of a CANN is decomposed into different modes
corresponding to distortions in 
the height, position, width or skewness of the network bump states.
Their contributions to the tracking performance
are determined by the amplitudes of the corresponding eigenvalues
of the neuronal interaction kernel.
For a CANN, neutral stability implies that the mode of
positional shift dominates the network dynamics 
and that other modes have high-order contributions.
From this property, we develop a time-dependent perturbative approach
to simplify the network dynamics.
Geometrically, this corresponds to projecting the network dynamics
on its dominating motion modes.
The solution of the perturbation method is expressed in a simple closed form,
and we can approximate the network dynamics
up to an arbitrary accuracy depending on the order of perturbation used.
With this method, we explore two tracking performances of a CANN,
namely, the condition under which the network can successfully track
a moving stimulus and the reaction time for the network
to catch up with sudden changes in the stimulus.
Both properties are associated with the
unique dynamics of a CANN and can be tested in practice.

The idea of utilizing the special structure of a CANN
to simplify its dynamics has also been reported in other studies.
Wu {\it et al} projected the dynamics of a CANN
on its single dominating motion mode,
the tangent of the attractor space,
and simplify the network dynamics into a one-dimensional equation~\cite{Wu08}.
Several other authors also proposed different approaches
to simplifying the network dynamics
~\cite{Ben-Yishai95,Wennekers02,Renart03,Miller06}.
Compared with these studies, our contributions are that:
1) we elucidate the tracking behavior of a CANN in much more detail,
particularly the effect of distortions in the network state
on the tracking performance and the development of a perturbative method
to systematically improve the description of the network dynamics;
2) we predict on the tracking behavior of a CANN, namely,
the maximum speed for a moving stimulus to be trackable by the network
\cite{Hansel1998} and the logarithmic nature of the reaction time; and
3) we also analytically investigate the tracking dynamics
of a two-dimensional CANN.
We expect that the mathematical framework developed in this work
will have wide applications in the theoretical study of CANNs.

The organization of the paper is as follows.
In Section~2, the unique dynamics of a CANN is explored,
and we demonstrate how a CANN is decomposed into different modes.
In Section~3, a simple method of projecting the network dynamics
onto its single primary motion mode,
i.e., the tangent of the manifold formed by all stationary states,
is used to investigate the tracking performance of a CANN.
In Section~4, a perturbative approach is developed
to systematically improve the description
of the tracking dynamics of the network.
The effects of the shape distortions of the network state
on the tracking dynamics are described.
In Section~5, the dynamics of a two-dimensional CANN is investigated.
In Section~6, we present the discussion and conclusions. 
The mathematical details are presented in the appendices.
Preliminary results have been published in~\cite{Fung08}.

\begin{figure*}[htb]
\begin{center}
\epsfig{file=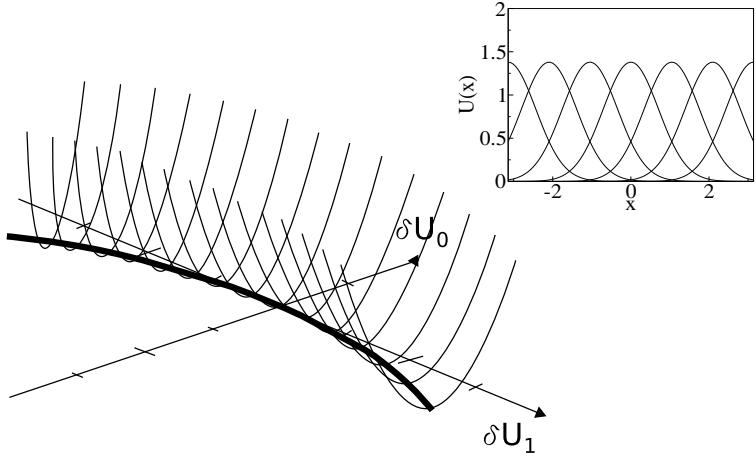,width=10cm}
\caption{The energy landscape of the state space of a CANN.
A canyon is formed by projecting the stationary states
onto the subspace formed by $\delta U_1$ and $\delta U_0$,
i.e., the position and the height distortions of the network states.
Motion along the canyon corresponds to the positional shift of the
bump (inset).}
\end{center}
\end{figure*}

\section{Intrinsic Dynamics of a CANN}

Let us consider a one-dimensional continuous stimulus $x$
encoded by an ensemble of neurons.
For example, the stimulus may represent the direction of movement,
the orientation or a general continuous feature of objects
extracted by the neural system. 
We consider the range of possible values of the stimulus
being much larger than the range of neuronal interactions. 
We can thus effectively take $x\in(-\infty,\infty)$.

Let $U(x,t)$ be the synaptic input at time $t$
to the neurons whose preferred stimulus is $x$, 
and let $r(x,t)$ be the firing rate of these neurons.
$r(x,t)$ increases with the synaptic input,
but saturates in the presence of global activity-dependent inhibition.
A solvable model that captures these features is given 
by the divisive normalization~\cite{Deneve99,Wu02}:
\begin{equation}
    r(x,t)=\frac{U(x,t)^2}{1+k\rho\int_{-\infty}^{-\infty} dx'U(x',t)^2},
\label{eq:output}
\end{equation}
where $\rho$ is the neural density 
and $k$ is a small positive constant
controlling the strength of global inhibition.

The dynamics of the synaptic input $U(x,t)$
is determined by the external input $I_{\rm ext}(x,t)$,
the network input from other neurons, and its own relaxation.
It is given by
\begin{equation}
    \tau\frac{\partial U(x,t)}{\partial t}
    =I_{\rm ext}(x,t)+\rho\int^\infty_{-\infty} dx'J(x,x')r(x',t)-U(x,t),
\label{eq:dyn}
\end{equation}
where $\tau$ is a time constant,
which is typically on the order of $1$ ms~\cite{Ermentrout98,Gutkin03,Wu08},
and $J(x,x')$ is the neural interaction from $x'$ to $x$.

The key character of CANNs is
the translational invariance of their neural interactions.
In our solvable model,
we choose Gaussian interactions with a range $a$, namely,
\begin{equation}
    J(x,x')=\frac{A}{\sqrt{2\pi}a}
    \exp\left[-\frac{(x-x')^2}{2a^2}\right],
\label{eq:interactions}
\end{equation}
where $A$ is a constant.

CANNs with other neural interactions and inhibition mechanisms have
been studied in the literature (see, e.g.,
\cite{Amari77,Ben-Yishai95,Ermentrout98,Wang01}). 
Here, we choose
this model for the advantage that it permits an analytical solution
of the network dynamics. Nevertheless, the final conclusions of our
model are qualitatively applicable to general cases.

We first consider the intrinsic dynamics of the CANN model in the
absence of external stimuli. 
It is straightforward to check that, 
for $0<k<k_c\equiv A^2\rho/(8\sqrt{2\pi}a)$, the network holds a
continuous family of stationary states, which are (see Fig.~2A)
\begin{eqnarray}
    \tilde U(x|z) & = & U_0\exp\left[-\frac{(x-z)^2}{4a^2}\right],
    \label{eq2.4}
    \\
    \tilde r(x|z) & = & r_0\exp\left[-\frac{(x-z)^2}{2a^2}\right],
\label{eq2.5}
\end{eqnarray}
where $U_0=[1+(1-k/k_c)^{1/2}]A/(4\sqrt\pi ak)$ and 
$r_0=[1+(1-k/k_c)^{1/2}]/(2\sqrt{2\pi}ak\rho)$. These stationary
states are translationally invariant among themselves and have a 
Gaussian shape with a free parameter $z$ indicating their positions.
$z$ is therefore the coordinate of the manifold formed by all
stationary states.

The stability of the Gaussian bumps can be studied
by considering the dynamics of fluctuations.
Consider the network state $U(x,t)=\tilde U(x|z)+\delta U(x,t)$.
As derived in Appendix B, we have
\begin{equation}
    \tau\frac{\partial}{\partial t}\delta U(x,t)
    =\int^\infty_{-\infty} dx'F(x,x'|z)\delta U(x',t)
    -\delta U(x,t),
\label{eq:fluc}
\end{equation}
where the interaction kernel $F(x,x'|z)$ is given by
\begin{equation}
    F(x,x'|z)=\frac{2\rho U(x')}{B}
    \left[J(x,x')-k\rho\int^\infty_{-\infty} dx'' J(x,x'')r(x'')\right].
\label{eq:kernel}
\end{equation}

We are interested in the eigenfunctions and eigenvalues
of the kernel $F(x,x'|z)$. 
Since the kernel is not symmetric 
with respect to the permutation of $x$ and $x'$, 
we have to distinguish between its left and right eigenfunctions. 
To compute them, 
we need to choose a convenient set of basis functions.
They should satisfy the boundary condition
that they vanish when $x$ approaches $\pm\infty$.
A convenient choice is one that contains a basis function
taking the shape of a Gaussian bump.
Since the quantum ground state of the harmonic oscillator potential
has a Gaussian shape,
a natural choice of basis functions is the wave functions
of quantum harmonic oscillators~\cite{Griffiths04}, namely,
\begin{equation}
    v_n(x|z)=\frac{1}{\sqrt{(2\pi)^{1/2}an!2^n}}
    \exp\left[-\frac{(x-z)^2}{4a^2}\right]
    H_n\left(\frac{x-z}{\sqrt 2a}\right),
\label{eq:basis}
\end{equation}
where $H_n$ are the $n$th-order Hermite polynomials
for $n=0,1,2,\ldots$,
which can be expressed
as the derivatives of the Gaussian function~\cite{Griffiths04}
\begin{equation}
    v_n(x|z)={\frac {(-1)^n(\sqrt{2}a)^{n-1/2}} {\sqrt{\pi^{1/2} n! 2^n}}}
    \exp\left[{\frac {(x-z)^2} {4a^2}}\right]\left(\frac{d}{dx}\right)^n
    \exp\left[-{\frac {(x-z)^2} {2a^2}}\right].
\end{equation}
This choice of basis functions has the further advantage
that an abundance of mathematical properties
are available for our analysis,
as summarized in Appendix A.

Fig.~2B pictures the first four wave functions. We see that:
\begin{itemize}
\item The ground state of the quantum harmonic oscillator,
\begin{equation}
    v_0(x|z)=\frac{\tilde{U}(x|z)}{U_0\sqrt{(2\pi)^{1/2}a}},
\label{eq2.9}
\end{equation}
corresponds to the stationary state of the network.
\item The first excited state of the quantum harmonic oscillator,
\begin{equation}
    v_1(x|z)=\frac{2a}{U_0\sqrt{(2\pi)^{1/2}a}}
    \frac{\partial \tilde{U}(x|z)}{\partial z},
\label{eq2.10}
\end{equation}
corresponds to the positional shift of the stationary state.
\item
The second and third excited states, $v_2(x|z)$ and $v_3(x|z)$, 
respectively correspond to the distortions in the width and skewness
of the bump. 
\end{itemize}

\vskip 0.5cm
\begin{figure*}[htb]
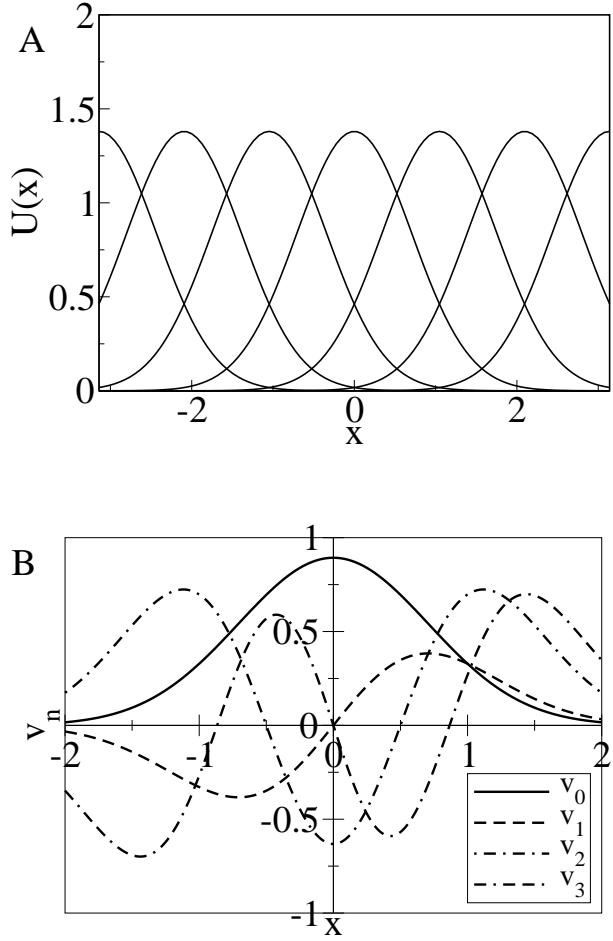

\begin{center}
\epsfig{file=fig2a.eps,width=8cm}
\\
\vskip 1cm
\epsfig{file=fig2b.eps,width=8cm}
\caption{(A) The Gaussian-shaped stationary states of the network.
(B) The first four basis functions of the quantum harmonic oscillators,
representing distortions in 
the height, position, width and skewness of the Gaussian bump.}
\end{center}
\end{figure*}

To express $\delta U(x,t)$ in terms of the eigenfunctions
$v_n(x|z)$ using Eq.~(6),
we consider $F(x,x'|z)$ as an operator acting on $v_n(x|z)$.
Since $v_n(x|z)$ is a basis, we have
\begin{equation}
    \int^\infty_{-\infty} dx'F(x,x'|z)v_n(x'|z)
    =\sum_m v_m(x|z)F_{mn},
\label{eq2.11}
\end{equation}
where $F_{mn}$ is the representation of the kernel $F(x,x'|z)$
in the basis $v_n(x|z)$, given by
\begin{equation}
    F_{mn}=\int^\infty_{-\infty}dx\int^\infty_{-\infty}dx'
    v_m(x|z)F(x,x'|z)v_n(x'|z).
\label{eq:fmn_def}
\end{equation}
As derived in Appendix B for the particular choice
of the output function (\ref{eq:output})
and the neural interactions (\ref{eq:interactions}),
the elements of the matrix $F_{mn}$ are given by
\begin{equation}
    F_{mn}=\left\{\begin{array}{ll}
    1-\sqrt{1-k/k_c}, & m=n=0;\\
    2^{1-n}\sqrt{\frac{n!}{m!}}\frac{(-1)^{\frac{n-m}{2}}}
    {2^{\frac{n-m}{2}}\displaystyle\left(\frac{n-m}{2}\right)!},&
    \mbox{$n-m$ being an even integer};\\
    0, & \mbox{otherwise}.
    \end{array}\right.
\label{eq:fmn}
\end{equation}

By using the matrix $F_{mn}$,
we can calculate the eigenvalues and
the first four right eigenfunctions of the kernel $F(x,x'|z)$, which
are (see Appendix B)
\begin{eqnarray}
    \lambda_0 & = & 1-\sqrt{1-k/k_c},
\label{eq:l0} \\
    \lambda_n & = & 2^{1-n}, ~~~~\mbox{for}~~n\geq 1,\ldots,
\label{eq:ln} \\
    u^R_0(x|z) & = & v_0(x|z),
\label{eq:u0} \\
    u^R_1(x|z) & = & v_1(x|z),
\label{eq:u1} \\
    u^R_2(x|z) & = & {\frac {\sqrt{1/2}} {D_0}}v_0(x|z)
    +\frac{1-2\sqrt{1-k/k_c}}{D_0}v_2(x|z),
\label{eq:u2} \\
    u^R_3(x|z) & = & \sqrt{\frac{1}{7}}v_1(x,z)
    + \sqrt{\frac{6}{7}}v_3(x,z),
\label{eq:u3}
\end{eqnarray}
where
\begin{equation}
    D_0=\left[\left(1-2\sqrt{1-k/k_c}\right)^2+1/2\right]^{1/2}.
\label{eq:d0}
\end{equation}
We note that $u^R_1(x|z)=v_1(x|z)$ and $\lambda_1=1$.
Fig.~3A illustrates the first four right eigenfunctions of $F$.

The right eigenfunctions of $F$ correspond to the various distortion
modes of the bump.
For example,
$u^R_0(x|z)$ corresponds to the amplitude distortion
of the bump,
and $u^R_1(x|z)$ to the positional shift.
Starting from $n=3$, $u^R_n(x|z)$ are linear combinations
of $v_k(x|z)$, where $k=n, n-2,\ldots$.
Some of them may be dominated by a specific type of distortion,
such as $u^R_3$ in Eq.~(\ref{eq:u3})
being dominated by the skewness.
However, generally speaking, various distortion features
have been coupled in the eigenfunctions of the interaction kernel.
To avoid confusion,
we continue to use $v_n(x|z)$ as our basis functions
in the rest of the paper.

To identify the contributions of $u^R_n(x|z)$ to the network dynamics,
we express $\delta U(x,t)$ in (\ref{eq:fluc})
as a linear combination of $u^R_n(x|z)$, namely,
$\delta U(x,t)=\sum_n\delta U_n(z,t)u^R_n(x|z)$.
Using the orthonormality of the left and right eigenfunctions
of $F(x,x'|z)$, we have
\begin{equation}
    \delta U_n(z,t)=\int^\infty_{-\infty}dx\delta U(x,t)u^L_n(x|z).
\end{equation}
Assume that the motion of the bump is slow,
so that $dz/dt$ becomes negligible in Eq.~(\ref{eq:fluc});
as we shall see,
this assumption is valid as long as the external stimulus is sufficiently weak.
Then, the projection of Eq.~(\ref{eq:fluc}) on the eigenfunctions become
\begin{equation}
    \tau\frac{d}{dt}\delta U_n(z,t)=(\lambda_n-1)\delta U_n(z,t).
\label{eq2.19}
\end{equation}
Hence,
\begin{equation}
    \delta U_n(z,t)=\delta U_n(z,0)
    \exp\left[-{\frac {(1-\lambda_n)t} {\tau}}\right],
\label{eq2.20}
\end{equation}
where $\delta U_n(z,0)$ is the initial value of the projection.

Eq.~(\ref{eq2.20}) tells us that, since $\lambda_1=1$,
the bump distortion $\delta U_1$,
which corresponds to the positional shift of the bump,
is sustained under the network dynamics. This is the mathematical expression
of the neutral stability of the stationary states of the network.
The eigenvalues for all other eigenfunctions are all less than one.
Since these other modes of bump distortions decay exponentially
with the time constant $\tau/(1-\lambda_n)$,
their contributions on the network dynamics are
small when compared with that due to the positional shift.
An intuitive illustration of the decomposition of the network dynamics
is given in Fig.~3B.

\vskip 0.5cm
\begin{figure*}[htb]
\begin{center}
\epsfig{file=fig3a.eps,width=8cm}
\\
\vskip 1cm
\epsfig{file=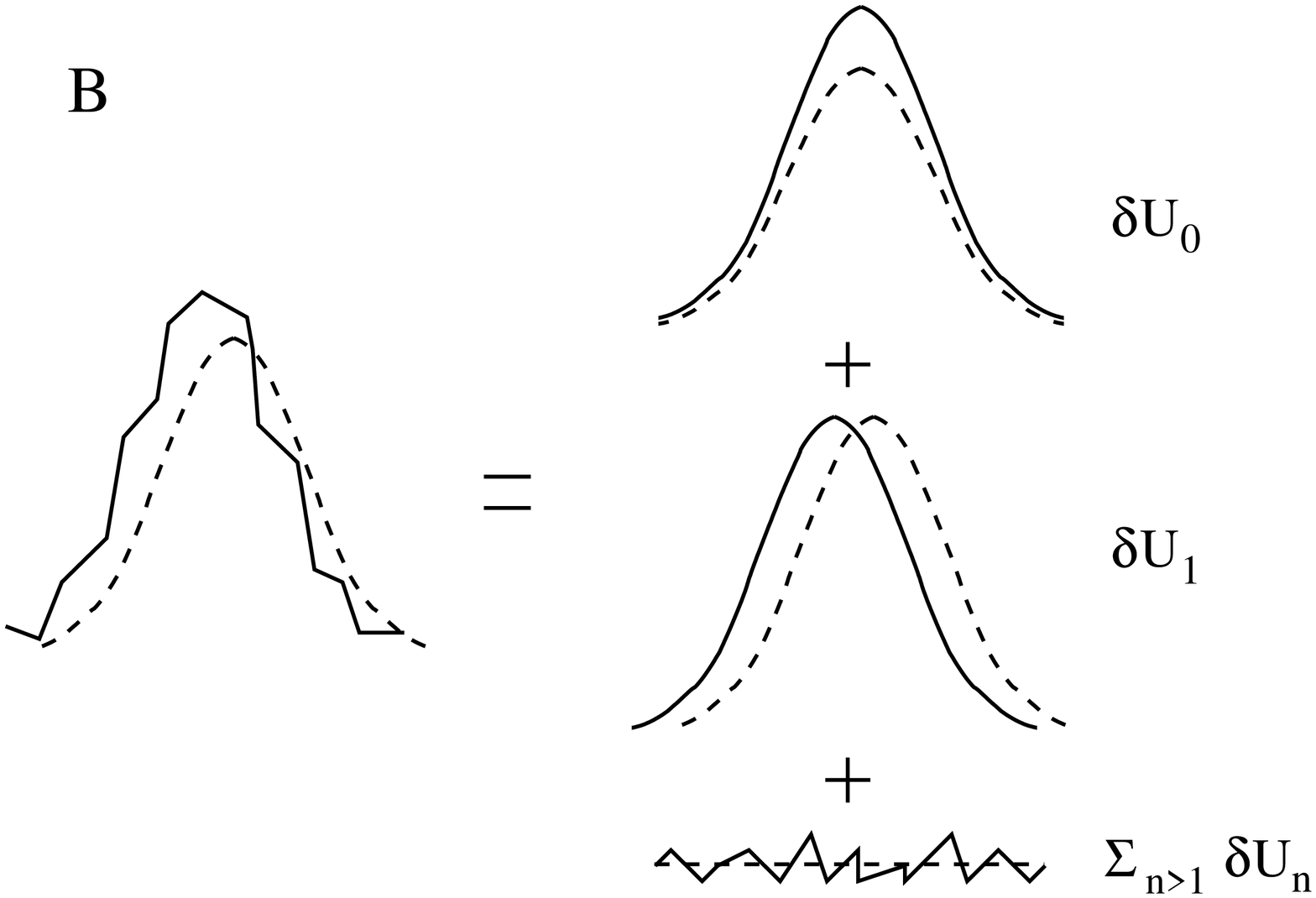,width=8cm}
\caption{(A) The first four right eigenfunctions of the interaction kernel
 $F(x,x'|z)$  with $k/k_c = 1/2$.
(B) The distortions of the network state are decomposed into three parts,
namely, distortions in the height, position and all other shapes.}
\end{center}
\end{figure*}

The fact that the dynamics of a CANN is dominated by only a few motion modes
suggests that we may use this property
to simplify the description of the network dynamics; 
that is, we can project the network dynamics onto its dominating modes.
By this, we can analytically quantify the tracking bahviors of a CANN.

\subsection*{The Landscape of the State Space}

It is instructive to illustrate the energy landscape 
in the state space of a CANN. 
In the dynamical system that we are studying, 
we do not have a Lyapunov function.
This is evident from the fact that the interaction matrix
$F_{mn}$ is not symmetric.
Nevertheless, at each peak position $z$,
one can define an effective energy function
$E(z)=\sum_n(1-\lambda_n)[\delta U_n(z)]^2/2$,
with $\delta U_n(z)$ being the projection of $U(x)-\tilde{U}(x|z)$
on the  eigenfunction $u_n(x|z)$.
Then, the dynamics in Eq.~(\ref{eq:fluc}) or (\ref{eq2.19})
can be locally described by the gradient descent of $E(z)$
in the space of $\delta U_n(z)$.
$E(z)$ has the minimum value of zero when
$\delta U_n(z)=0$ for $n\ne 1$
and $\delta U_1(z)$ can take any value since $\lambda_1=1$.
This corresponds to the picture that the bump's position shifts, 
but the shape of the bump remains unchanged.
When $z$ varies, the solution of $E(z)=0$
traces a curve in the state space,
facilitating the local gradient descent dynamics of the network,
with the tangent of the curve at each point
defining the direction of neutral stability.
We can envisage this by a canyon surrounding the curve
as shown in Fig.~1, 
and a small force along the tangent of the canyon
can move the network state easily.

\subsection*{The External Input}

To proceed, we need to define the external input.
Without loss of generality, we choose
\begin{equation}
    I_{\rm ext}(x,t)=\alpha U_0\exp\left[-{\frac {(x-z_0)^2} {4a^2}}\right]
    +\sigma \eta(x,t),
\label{eq:input}
\end{equation}
where $\alpha$ and $\sigma$ indicate 
the strength of the signal and the noise, respectively.
$\eta(x,t)$ is Gaussian white noise, satisfying
$\langle \eta(x,t) \rangle=0$ and
$\langle \eta(x,t)\eta(x',t') \rangle=\delta(x-x')\delta(t-t')$.
The signal term has the effect of driving the bump to
the stimulus location $z_0$,
and the noise term tends to shift the bump position randomly.
Our results are qualitatively independent of the exact form
of the external input.
We may choose other forms for the external input.
The main conclusions will be the same
once the chosen external input has a unimodal shape
and has the effect of driving the bump to the stimulus position.

\subsection*{The Setting of the Simulation Experiments}

We will use simulation experiments to confirm the theoretical analysis.
Here, we describe the setting of simulation experiments to be used below.
The network consists of $N$ neurons.
Their preferred stimuli $x$ are evenly distributed in the range
$(-\pi,\pi]$ and satisfy the periodic condition, namely,
$U(x,t)=U(x+2\pi,t)$.
In the simulation, the integration in the continuum limit
is changed to the summation accordingly,
i.e., $\rho \int dx\to\sum_{i=1}^{N}$. We choose $a=0.5\ll 2\pi$,
so that comparison with the theoretical analysis for
$x\in (-\infty,\infty)$ holds.

\section{Tracking Dynamics in the Weak Input Limit}

We consider first the case when the external input is sufficiently small, 
i.e., $\alpha\ll 1$ and $\sigma\ll 1$.
In this case, we can reasonably assume
that the network dynamics is dominated by the positional shift of the bump,
and we can thus ignore other motion modes, that is,
$U(x,t)\approx \tilde{U}(x|z(t))$, with only
the bump position $z(t)$ varying with time.
As will be shown in Section~4,
this actually corresponds to $n=0$ perturbation in
the general perturbative method.

After projecting the network dynamics, Eq.~(\ref{eq:dyn}),
on the basis functions $v_1(x|z(t))$, we obtain (see Appendix C)
\begin{equation}
    \frac{dz}{dt}=-\frac{\alpha}{\tau}(z-z_0)
    \exp\left[-\frac{(z-z_0)^2}{8a^2}\right]
    +\sqrt{2D}\eta_1(t),
\label{OU}
\end{equation}
where $D=\sqrt{2}a\sigma^2/(U_0^2\tau^2\sqrt{\pi})$ is the diffusion constant,
and $\eta_1(t)$ is the overlap of $\eta(x,t)$ with $v_1(x|z(t))$ given by
\begin{equation}
    \eta_1(t)=\int^\infty_{-\infty} dx\eta(x,t)v_1(x|z(t)),
\label{eq:noise1}
\end{equation}
which satisfies
\begin{equation}
    \langle\eta_1(t)\rangle=0
    \quad{\rm and}\quad
    \langle\eta_1(t)\eta_1(t')\rangle=\delta(t-t').
\label{eq:noiseav}
\end{equation}

This equation represents a one-dimensional Ornstein-Uhlenbeck process
for the motion of the bump position in the state space~\cite{Tuckwell88}.
Its meaning is straightforward:
the first term on the right-hand side of the equation
represents the contribution of the stimulus,
whose effect is to pull the bump to the stimulus position $z_0$;
the second term represents noise, 
which tends to shift the bump position randomly.
This equation agrees with Eq.~(3.7) in \cite{Wu08},
which considers $|z-z_0|\ll a$ and hence $e^{-(z-z_0)^2/8a^2}\approx 1$.

The distribution of the bump position can be calculated.
The pull term in Eq.~(\ref{OU}) can be written as the gradient of a potential,
\begin{equation}
    V(z)=-\frac{4a^2\alpha}{\tau}\exp\left[-\frac{(z-z_0)^2}{8a^2}\right],
\label{eq3.2}
\end{equation}
so that Eq.~(\ref{OU}) can be written as a Langevin equation,
\begin{equation}
    \frac{dz}{dt}=-\frac{\partial V}{\partial z}
    +\sqrt{2D}\eta(t).
\label{eq3.3}
\end{equation}
This implies that $z$ satisfies a Boltzmann distribution~\cite{Ma76},
\begin{equation}
    P(z)\sim \exp\left[-\frac{V(z)}{D}\right].
\label{boltz}
\end{equation}
Since the pull is maximum at $z=z_0\pm 2a$,
two noise regimes can be identified.
In the low noise regime, where $\langle (z-z_0)^2\rangle\ll 4a^2$,
the bump position is effectively trapped in a parabolic potential, 
whereas in the high noise regime with $\langle (z-z_0)^2\rangle\gg 4a^2$,
the pull vanishes and the motion of the bump resembles
a random walk with $\langle[z(t)-z_0]^2\rangle=2Dt$
loosely coupled to the stimulus.

However, simulation results show that Eq.~(\ref{boltz})
underestimates the variance of the bump position.
This is because we have considered only 
the overlap of $\eta(x,t)$ with $v_1(x|z(t))$.
As will be shown in the perturbative approach,
overlaps with other components cannot be neglected.

By using the simplified dynamics in Eq.~(\ref{OU}),
we explore two tracking behavior of the network,
namely, the tracking of a moving stimulus
and the catching up of an abrupt change in the stimulus. 

\subsection{The Condition for Successful Tracking}

Suppose that the stimulus is moving at a constant velocity $v$
and the noise is low and negligible.
The dynamical equation becomes identical to Eq.~(\ref{OU})
after the transients, with $z_0$ replaced by $vt$.
Denoting the lag of the bump behind the stimulus as $s=z_0-z$, we have
\begin{eqnarray}
    {\frac {ds} {dt}} & = &{\frac {dz_0} {dt}}-{\frac {dz} {dt}}
    \nonumber \\
    &=&v-\frac{\alpha s}{\tau}\exp\left[-\frac{s^2}{8a^2}\right]
    \nonumber \\
    & = & v-g(s).
\label{eq3.5}
\label{tracking}
\end{eqnarray}
The size of the lag $s$ is determined by two competing factors:
the first term represents the movement of the stimulus,
which tends to enlarge the lag;
and the second term represents the collective effects
of the neuronal recurrent interactions,
which tends to shorten the lag.
The tracking is achieved when these two forces match each other,
i.e., $v=g(s)$;
otherwise, $s$ diverges.

Fig.~4A shows that the function $g(s)$ is concave 
and has the maximum value of
$g_{\rm max}=2\alpha a/(\tau\sqrt{e})$ at $s=2a$. 
This means that if $v>g_{\rm max}$, 
the network is unable to track the stimulus.
Thus, $g_{\rm max}$ defines the maximum trackable speed
of a moving stimulus to the network.
Notably, $g_{\rm max}$ increases with the strength of the external signal
$\alpha$ and the range of neuronal recurrent interactions $a$.
This is reasonable
since it is the neuronal interactions that
induce the movement of the bump. $g_{\rm max}$
decreases with the time constant of the network $\tau$, as this
reflects the responsiveness of the network to external inputs.

On the other hand, for $v<g_{\rm max}$,
there is a stable and unstable fixed point,
denoted as $s_1$ and $s_2$ respectively, in Fig.~4A.
When the initial lag is less than $s_2$, it will converge to $s_1$.
Otherwise, the tracking of the stimulus will be lost. 
Intuitively, this means that 
the propagation of neuronal activities in the network 
needs to be quick enough 
to compensate for the initial lag behind the stimulus.
Figs.~4B and 4C show that the analytical results of Eq.~(\ref{tracking})
agree well with the simulation results.

\begin{figure*}[htb]
\begin{center}
\epsfig{file=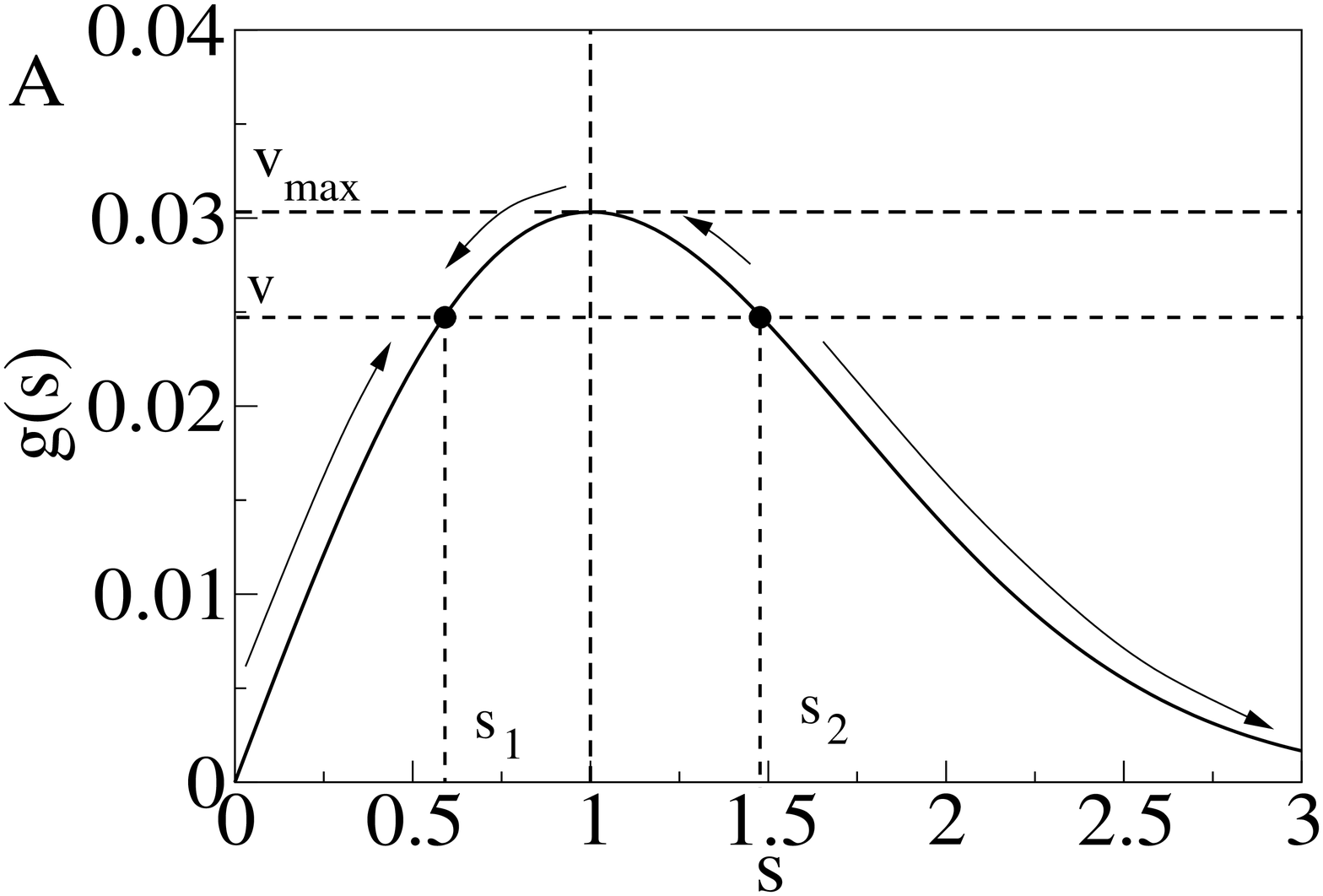,width=6cm}
\\
\vskip 1cm
\epsfig{file=fig4b.eps,width=6cm}
\\
\vskip 1cm
\epsfig{file=fig4c.eps,width=6cm}
\caption{(A) The function $g(s)$ and the stable and unstable
fixed points of Eq.~(\ref{eq3.5}). (B) The time dependence of the
separation $s$ starting from different initial values. $v=0.025$. (C)
The dependence of the terminal separation $s$ on the speed $v$.
Parameters: $N = 200$, $k = 0.5$, $a = 0.5$, $\tau = 1$,
$A = \sqrt{2\pi}a$, $\rho = N/2\pi$ and $\alpha = 0.05$.}
\end{center}
\end{figure*}

\subsection{The Reaction Time}

Another tracking behavior we study is the response to a stimulus
experiencing an abrupt change.
We measure how long it takes for the network to catch up with this change,
i.e., the reaction time. 
Suppose that the network has reached a steady state at $t<0$, and the stimulus
position jumps from $0$ to $z_0$ suddenly at $t=0$.
This is the typical setting in experiments studying mental
rotation behavior~\cite{Georgopoulos93,Ben-Yishai95,Wu05}.
The reaction time can be obtained by integrating
Eq.~(\ref{OU}) with the initial value $z(t=0)=0$.

When the noise is low and the jump size $|z_0|$
is much smaller than the range $a$ of neuronal interactions,
we can derive an analytical expression for the reaction time.
In this case, Eq.~(\ref{OU}) is approximated as
\begin{equation}
    \tau {\frac {dz} {dt}}\approx \alpha(z_0-z).
\label{eq3.6}
\end{equation}
We define the reaction time $T$ as the time it takes 
for the bump to move to a small threshold
distance $\theta$ from the stimulus. $T$ is calculated to be
\begin{equation}
    T \approx {\frac {\tau} {\alpha}}
    \ln \left({\frac {|z_0|} {\theta}}\right).
\label{eq3.7}
\end{equation}
This reveals that with small jump sizes, 
the reaction time of a CANN increases logarithmically with
the jump size.

Fig.~5 compares the simulation results
with our theoretical predictions based on Eq.~(\ref{OU}).
We see that with small jump sizes, they agree very well with each other, 
and the reaction time indeed has 
a logarithmic nature in its increase with the jump size.
However, we also notice that with large jump sizes,
there is considerable discrepancy between the simulation results 
and the theoretical predictions. 
This is because we have only considered
the positional shift in the network dynamics.
To improve the results, more motion modes must be included. 

\vskip 0.5cm
\begin{figure*}[htb]
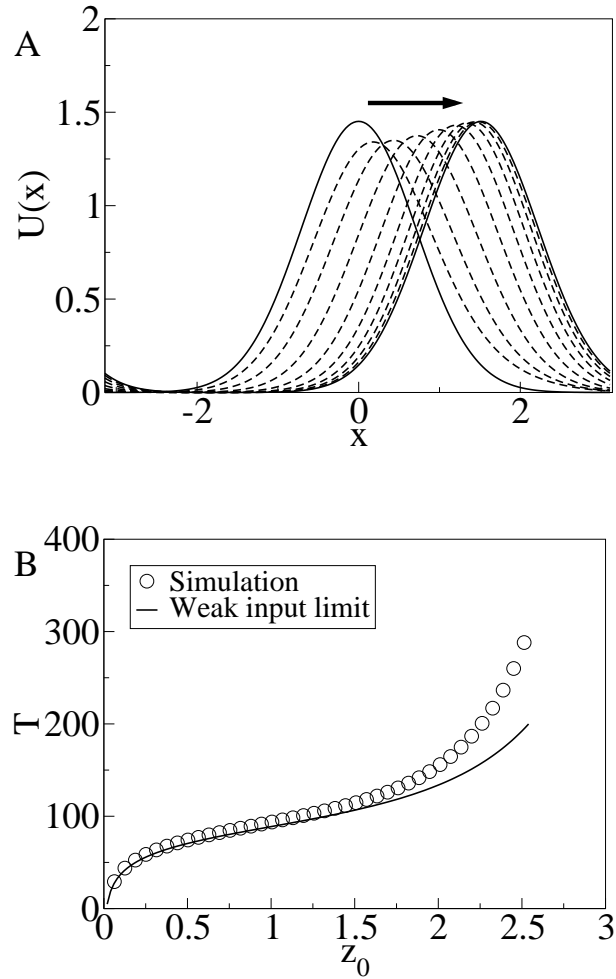

\begin{center}
\epsfig{file=fig5a.eps,width=8cm}
\\
\vskip 1cm
\epsfig{file=fig5b.eps,width=8cm}
\caption{(A) The tracking process of the network.
(B) The reaction time vs. the jump size $|z_0|$.
Parameters: $\theta = \pi/200$ and the rest are the same as Fig. 4.}
\end{center}
\end{figure*}

\section{Tracking Dynamics: A Comprehensive Study}

In this section, we develop a time-dependent
perturbation approach to systematically improve the description
of the network dynamics.
Since the network dynamics is primarily dominated
by the positional shift of the bump,
with high-order distortions from other motion modes,
we consider the network states with the following form,
\begin{equation}
    U(x,t)=\tilde{U}(x|z(t))+\sum_{n=0}^{\infty}a_{n}(t)v_n(x|z(t)),
\label{eq4.1}
\end{equation}
where the bump position $z(t)$ and the coefficients $\{a_n(t)\}$
specify the network state.

The external input can be similarly expressed as
\begin{equation}
    I_{\rm ext}(x,t)=\sum_{n=0}^\infty I_n(t)v_n(x|z),
\label{eq:modes}
\end{equation}
with $I_n=\int dx I_{\rm ext}(x,t)v_n(x|z)$.

By substituting Eqs.~(\ref{eq4.1}) and (\ref{eq:modes}) 
into Eq.~(\ref{eq:dyn}),
and making use of the orthonormality and completeness of $v_n(x|z)$,
we obtain the perturbative equation for $a_n(t)$ as (see Appendix D) 
\begin{eqnarray}
    \left(\frac{d}{dt}+\frac{1-\lambda_n}{\tau}\right)a_n
    & = &\frac{I_n}{\tau}
    -\left[ U_0\sqrt{(2\pi)^{1/2}a}\delta_{n1}+\sqrt{n}a_{n-1}
    -\sqrt{n+1}a_{n+1}\right]\frac{1}{2a}\frac{dz}{dt}
    \nonumber \\
    &+& \frac{1}{\tau}\sum_{r=1}^{\infty}\sqrt{\frac{(n+2r)!}{n!}}
    \frac{(-1)^r}{2^{n+3r-1}r!}a_{n+2r}.
\label{eq:perturb}
\end{eqnarray}

The center-of-mass position $z(t)$
is determined from the self-consistent condition, 
\begin{equation}
    z(t)=\frac{\int_{-\infty}^{\infty} dx U(x,t)x}
    {\int_{-\infty}^{\infty} dx U(x,t)}.
\label{eq:mass}
\end{equation}
By combining Eqs.(\ref{eq:perturb}) and (\ref{eq:mass}) 
and denoting the product of the positive integers 
$n,n-2,\ldots$ as $n!!$, 
we obtain an equation for the bump velocity (see Appendix D),
\begin{equation}
    \frac{dz}{dt}=\frac{2a}{\tau}\frac{I_1+\sum_{n=3,\rm odd}^{\infty}
    \sqrt{\frac{n!!}{(n-1)!!}}I_n+a_1}
    {U_0\sqrt{(2\pi)^{1/2}a}
    +\sum_{n=0,\rm even}^{\infty}\sqrt{\frac{(n-1)!!}{n!!}}a_n}.
\label{eq:velocity}
\end{equation}

Eqs.~(\ref{eq:perturb}) and (\ref{eq:velocity})
are the master equations of the perturbative method.
For the perturbation order specified by $n$,
we include the coefficients from $a_0$ up to $a_n$,
and the input components from $I_0$ up to $I_n$,
and at least $I_1$,
which is the lowest order of input that drives the dynamics.
By choosing different orders of perturbation,
we can approximate the network dynamics to different accuracies.

The components of the external input in $v_n(x|z)$ are calculated to be
(see Appendix E)
\begin{equation}
    I_n(t)=\alpha U_0 \exp\left[-\frac{(z_0-z)^2}{8a^2}\right]
    \left(\frac{z_0-z}{2a}\right)^n
    \sqrt{\frac{(2\pi)^{1/2}a}{n!}}+\sigma \eta_n(t),
\label{eq:stim_n}
\end{equation}
where the noise term $\eta_n(t)$ satisfies $\langle \eta_n(t)\rangle=0$
and $\langle \eta_m(t)\eta_n(t')\rangle=\delta_{mn}\delta(t-t')$.

\subsection{The Weak Input Limit}

We first compare the perturbative approach with the weak input limit 
by ignoring all distortion modes of the bump; 
that is, we set $a_n=0$ for all $n$.
Due to the readiness of the bump to shift its position,
the lowest order of the external stimulus to be included is $I_1$ 
according to Eq.~(\ref{eq:velocity}).
Thus, in the $n=0$ perturbation, Eq.~(\ref{eq:velocity}) becomes
\begin{equation}
    \frac{dz}{dt}=\frac{2a}{\tau}
    \frac{I_1}{U_0\sqrt{(2\pi)^{1/2}a}}.
\label{eq:v_weak}
\end{equation}
After substituting $I_1$ from Eq.~(\ref{eq:stim_n}) 
into Eq.~(\ref{eq:v_weak}),
we obtain Eq.~(\ref{OU}),
which is equivalent to the network dynamics in the limit of weak inputs.

\subsection{The $n=0$ Perturbation}

We further consider the $n=0$ perturbation.
This amounts to taking into account 
the effects of the change in the bump height on the network dynamics, 
apart from the positional shift. 
Perturbation up to $n=0$ yields
\begin{eqnarray}
    \left(\frac{d}{dt}+\frac{1-\lambda_0}{\tau}\right)a_0
    &=& \frac{\alpha U_0}{\tau}\sqrt{(2\pi)^{1/2}a}
    \exp\left[-\frac{(z_0-z)^2}{8a^2}\right]+\frac{\eta_0}{\tau},
\label{1st-order-a0}\\
    \frac{dz}{dt} &= & \frac{\alpha U_0\sqrt{(2\pi)^{1/2}a}
    (z_0-z)\exp\left[-\frac{(z_0-z)^2}{8a^2}\right]+2a\sigma\eta_1}
    {\tau (U_0\sqrt{(2\pi)^{1/2}a}+a_0)}.
\label{eq4.7}
\end{eqnarray}
With the initial condition $a_0=0$ at $t=0$,
the solution of $a_0$ from Eq.~({\ref{1st-order-a0}) reduces to
\begin{equation}
    a_0(t)=\int_0^{t}\frac{dt'}{\tau}
    \exp\left[-\frac{1-\lambda_0}{\tau}(t-t')\right]
    \left\{\alpha U_0\sqrt{(2\pi)^{1/2}a}
    \exp\left[-\frac{(z_0-z)^2}{8a^2}\right]
    +\eta_0(t')\right\}.
\label{eq4.8}
\end{equation}
Substituting Eq.~(\ref{eq4.8}) into Eq.~(\ref{eq4.7}), we have 
\begin{equation}
    \frac{dz}{dt}=\left\{\frac{\alpha}{\tau}(z_0-z)\exp\left[
    -\frac{(z_0-z)^2}{8a^2}\right]+\sqrt{2D}\eta_1 \right\}R(t)^{-1},
\label{eq:velocity1st}
\end{equation}
where
\begin{equation}
    R(t)=1+\alpha\int_0^t \frac{dt'}{\tau}\exp\left[
    -\frac{1-\lambda_0}{\tau}(t-t')-\frac{(z_0-z(t'))^2}{8a^2}\right]
    +\frac{\xi_0(t)}{U_0\sqrt{(2\pi)^{1/2}a}},
\label{eq4.10}
\end{equation}
and $\xi_0(t)$ is a stochastic variable given by
\begin{equation}
    \xi_0(t)=\sigma \int_0^t \frac{dt'}{\tau}
    \exp\left[-\frac{(1-\lambda_0)(t-t')}{\tau}\right]\eta_0(t').
\label{eq4.11}
\end{equation}

$R(t)$ can be interpreted to be the ratio of the bump height
in the presence of the external stimulus to its weak input limit.
Compared with Eq.~(\ref{OU}), Eq.~(\ref{eq:velocity1st})
has an extra term $R(t)^{-1}$.
$R(t)^{-1}$ approaches 1 in the limit of weak inputs, but, 
in general, $R(t)^{-1}<1$, implying that
the increase in the amplitude of the bump 
has the effect of slowing down the tracking dynamics.
In the following two applications,
we consider the low noise limit
with $\sigma=0$.

\subsubsection{Tracking a Moving Stimulus}

Similiar to the derivation of Eq.~(\ref{eq3.5}), 
we can determine 
the dynamical equation for the bump lag $s$ after the transient period.
From Eq.~(\ref{eq:velocity1st}),
\begin{equation}
    \frac{ds}{dt}=v-\frac{\alpha s}{\tau}
    \exp\left(-\frac{s^2}{8a^2}\right)R(t)^{-1},
\label{eq4.12}
\end{equation}
where $R(t)$ is given by
\begin{equation}
    R(t)=1+\alpha\int_{-\infty}^t \frac{dt'}{\tau}\exp\left[
    -\frac{1-\lambda_0}{\tau}(t-t')-\frac{s(t')^2}{8a^2}\right].
\end{equation}
If the stimulus is not too weak,
$s$ reaches its steady state value in a short duration.
The integral can then be computed precisely, yielding
\begin{equation}
    R(t)=1+\frac{\alpha}{1-\lambda_0}\exp\left[-\frac{s^2}{8a^2}\right].
\label{eq4.13}
\end{equation}
Compared with Eq.~(\ref{eq3.5}), Eq.~(\ref{eq4.12})
has a correction term, $R(t)^{-1}<1$, representing
the effect of the bump height distortion.

\subsubsection{Tracking an Abrupt Change in the Stimulus}

The dynamical equation in this case
is identical to Eq.~(\ref{eq:velocity1st})
with the initial value $z(t=0)=0$. To compute
the correction term $R(t)^{-1}$, however, we need to consider that
the bump height is initially at a steady value
determined by the steady-state response
of the network to the stimulus at $z=0$.
This means that Eq.~(\ref{eq4.8}) needs to be replaced by
\begin{eqnarray}
    a_0(t)& = & \frac{\alpha U_0\sqrt{(2\pi)^{1/2}a}}{1-\lambda_0}
    \exp\left[-\frac{(1-\lambda_0)t}{\tau}\right] \nonumber \\
    &+& \alpha U_0\sqrt{(2\pi)^{1/2}a}\int_0^{t}\frac{dt'}{\tau}
    \exp\left[-\frac{1-\lambda_0}{\tau}(t-t')-\frac{(z_0-z)^2}{8a^2}
    \right],
\label{eq4.14}
\end{eqnarray}
and the factor $R(t)$ in Eq.~(\ref{eq:velocity1st}) is replaced by
\begin{equation}
    R(t)=1+\frac{\alpha}{1-\lambda_0}
    \exp\left[-\frac{(1-\lambda_0)t}{\tau}\right]+
    \alpha\int_0^t \frac{dt'}{\tau}\exp\left[
    -\frac{1-\lambda_0}{\tau}(t-t')-\frac{(z_0-z(t'))^2}{8a^2}\right].
\label{eq4.15}
\end{equation}
Indeed, $R(t)$ represents the change in height during the movement of bump.
The second and third terms respectively contribute 
to the heights near the initial and final positions (see Fig.~5A).

Fig.~6 compares the simulation results and the theoretical predictions
based on the perturbation approximation
of Eqs.~(\ref{eq:velocity1st}) and (\ref{eq4.15}).
We see that compared with the weak input limit,
the result is improved a little bit.

\vskip 0.5cm
\begin{figure*}[htb]
\begin{center}
\epsfig{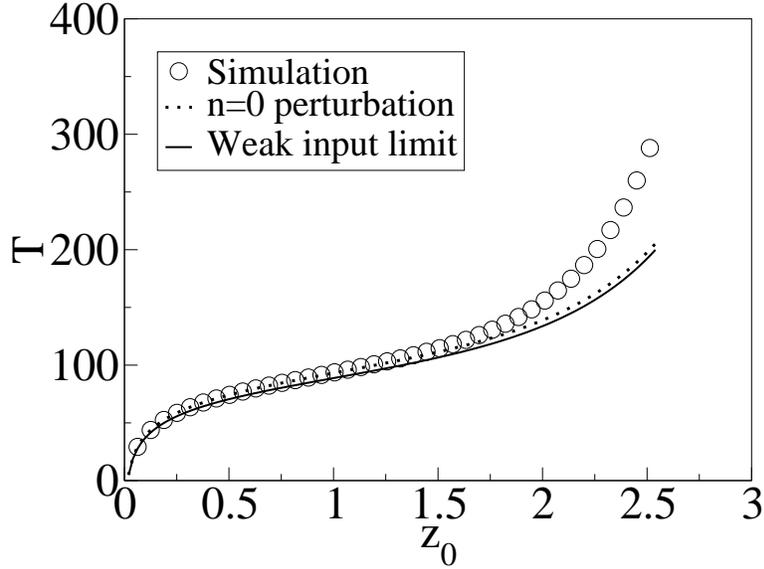}
\caption{The reaction time vs. the jump size.
 Parameters: the same as Fig. 5.}
\end{center}
\end{figure*}

\subsection{The $n=5$ Perturbation}

By increasing the order of perturbation,
we get an increasingly better approximation of the network dynamics.
We observe that the improvement is btter 
when the order of perturbation $n$ increases by 1 to an odd integer.
This is because odd distortion modes are antisymmetric, 
which better approximate the bump's shape distortion 
when the position of the external stimulus
is biased towards one side of the bump.

Below, we present the results of the $n=5$ perturbation,
which correspond to taking into account distortions in the position,
height, width, skewness and the fourth- and fifth-order distortions 
of the bump.

By including perturbations up to $n=5$,
Eqs.~(\ref{eq:perturb}) and (\ref{eq:velocity}) become
\begin{eqnarray}
    && \left(\frac{d}{dt}+\frac{1-\lambda_0}{\tau}\right)a_0=
    \frac{I_0}{\tau}+\frac{a_1}{2a}\frac{dz}{dt}
    -\frac{\sqrt{2}}{4\tau}a_2+\frac{\sqrt{6}}{32\tau}a_4,
\label{eq4.16} \\
    && \frac{da_1}{dt}=\frac{I_1}{\tau}-\left(\sqrt{(2\pi)^{1/2}a}U_0
    +a_0-\sqrt{2}a_2\right)\frac{1}{2a}\frac{dz}{dt}
    -\frac{\sqrt{6}}{8\tau}a_3+\frac{\sqrt{30}}{64\tau}a_5,
\label{eq4.17} \\
    && \left(\frac{d}{dt}+\frac{1}{2\tau}\right)a_2=
    \frac{I_2}{\tau}-\frac{\sqrt{2}a_1-\sqrt{3}a_3}{2a}
    \frac{dz}{dt}-\frac{\sqrt{3}}{8\tau}a_4,
\label{eq4.18} \\
    && \left(\frac{d}{dt}+\frac{3}{4\tau}\right)a_3=
    \frac{I_3}{\tau}-\frac{\sqrt{3}a_2-\sqrt{4}a_4}{2a}
    \frac{dz}{dt}-\frac{\sqrt{5}}{16\tau}a_5,
\label{eq4.19} \\
    && \left(\frac{d}{dt}+\frac{7}{8\tau}\right)a_4=
    \frac{I_4}{\tau}-\frac{\sqrt{4}a_3-\sqrt{5}a_5}{2a}
    \frac{dz}{dt},
\label{eq4.20} \\
    && \frac{dz}{dt}=\frac{2a\left(
    I_1+\sqrt{\frac{3}{2}}I_3+\sqrt{\frac{15}{8}}I_5+a_1\right)}
    {\tau\left(\sqrt{(2\pi)^{1/2}a}U_0+a_0+\sqrt{\frac{1}{2}}a_2+
    \sqrt{\frac{3}{8}}a_4\right)},
\label{eq4.21}
\end{eqnarray}
and $a_5$ is determined by the condition
$a_1+\sqrt{3/2}a_3+\sqrt{15/8}a_5=0$.
Fig.~7 compares the theoretical predictions with the simulation results.
They agree very well even when the jump sizes are large. 

\vskip 0.5cm
\begin{figure*}[htb]
\begin{center}
\epsfig{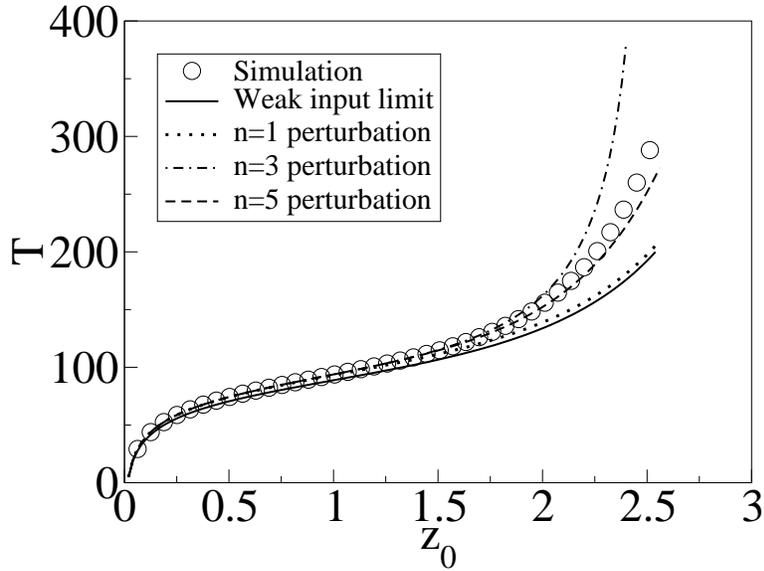}
\caption{The reaction time vs. the jump size. 
The simulation result is
compared with the $n=5$ perturbation approximation.
Parameters: the same as Fig.~5.  
Note that the predictions of the $n=0$ and $n=1$ perturbations are the same 
because of the self-consistent condition in Eq.~(\ref{eq:cm}). 
Also, the predicted reaction time given by the $n=1$ and $n=2$ perturbations 
are almost the same. 
So are the $n=3$ and $n=4$ perturbations.}
\end{center}
\end{figure*}

\subsection{Shape Distortion}

Besides the reaction time,
it is instructive to study the distortion in the bump's shape
during tracking. 
As shown in Fig. 8,
there is an abrupt change in the height, position, width and skewness
immediately after the external stimulus is abruptly removed at $t=0$.
$a_0(t)$ approaches 0 roughly,
as if there were no external stimulus,
and there is even a slight overshoot.
$a_1(t)$ also drops abruptly,
showing that the peak position lags behind the center of mass,
an indication that the bump is abruptly distorted.
At the same time, 
distortions in the width and skewness increase abruptly,
showing that the bump is suddenly pulled
by the external stimulus
from its new position.
After the initial abrupt changes,
the distorted components undergo rather smooth changes
on the journey to the new position.
Finally, when the bump approaches its destination,
the height returns to its equilibrium value,
while the other components approach 0.

We also compare the perturbation and simulation results
in Fig.~8.
The agreement is very good. 
Discrepancies arise from linearizing the interaction term 
in going from Eq.~(\ref{eq:dyn}) to (\ref{eq:fluc}). 
Their effects are mainly seen in the moments
immediately after the abrupt change in the external stimulus. 

Figure 9 shows the distorted bump
in the center-of-mass frame during tracking.
During the process, the lowering of the height,
the retarding of the peak position,
the broadening of the width,
and the skewing due to
the attraction of the newly positioned stimulus
are clearly visible.

\vskip 0.5cm
\begin{figure*}[htb]
\begin{center}
\epsfig{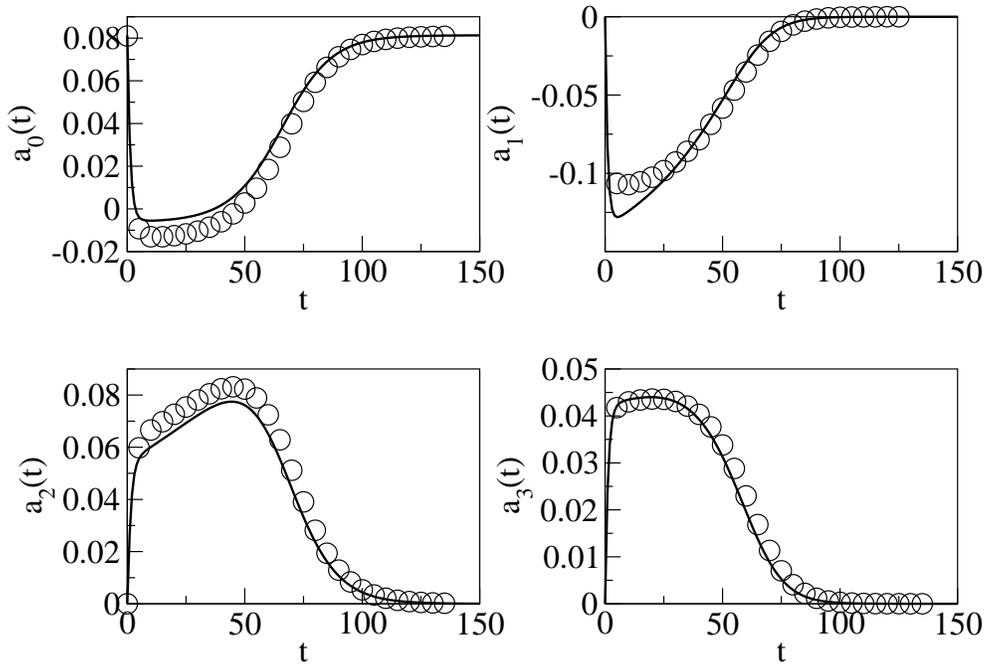}
\caption{
The first four $a_n(t)$
obtained by the $n=10$ perturbation (dashed lines)
and by projecting $U(x,t)$ onto the corresponding $v_n(x|z)$
during tracking (symbols).
Parameters: $N=200$, $\alpha=0.05$, $a=0.5$,
$\tau=1$, $k=0.5$, $\rho=N/2\pi$ and $A=\sqrt{2a}\pi$.}
\end{center}
\end{figure*}

\vskip 0.5cm
\begin{figure*}[htb]
\begin{center}
\epsfig{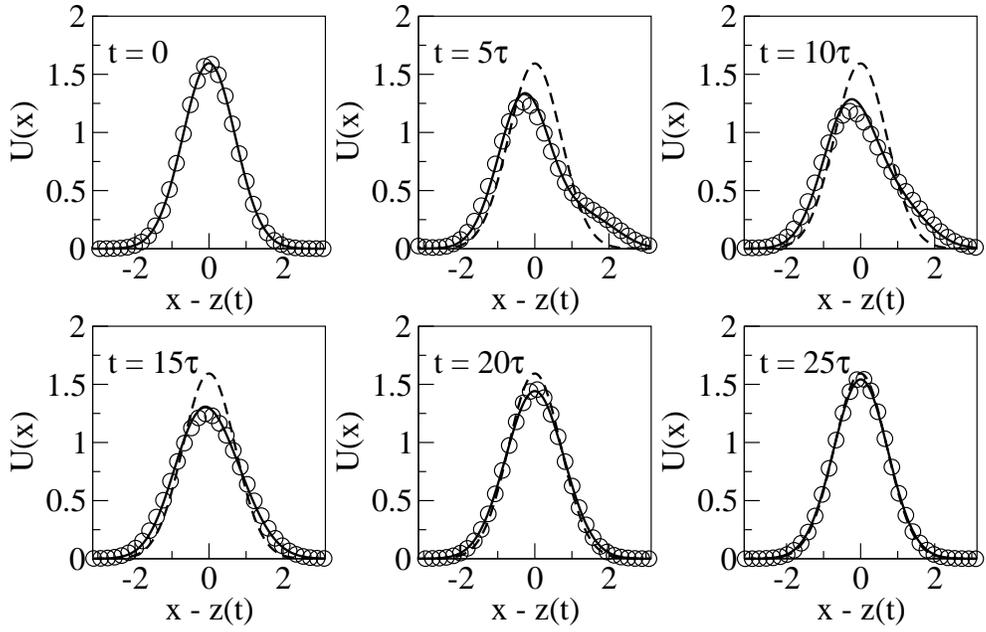}
\caption{
Snapshots of the neuronal inputs
in the center-of-mass frame of the bump.
Symbols: simulations.
Solid lines: predicted synaptic input
with the $n=10$ perturbation.
Parameters: $N=200$, $\alpha=0.15$, $a=0.5$,
$\tau=1$, $k=0.5$, $\rho=N/2\pi$ and $A=\sqrt{2a}\pi$.}
\end{center}
\end{figure*}

\section{Dynamics of a Two-Dimensional CANN}

We can extend our analysis to a two-dimensional(2D) CANN.
Consider a neural ensemble encoding a 2D continuous stimulus $\vx=(x_1,x_2)$,
with its components $x_1\in (-\infty,\infty)$
and $x_2\in(-\infty,\infty)$.
We denote by $U(\vx)$ the synaptic input to
neurons having the preferred stimulus $\vx$ and $r(\vx)$
the corresponding firing rate.
For a solvable model, we consider the network
whose dynamics is governed by
\begin{eqnarray}
    \tau\frac{\partial U(\vx,t)}{\partial t}
    &=&I_{\rm ext}(\vx,t)+\rho\int^\infty_{-\infty}\int^\infty_{-\infty}
    d\vx'J(\vx,\vx')r(\vx',t)-U(\vx',t),
\label{eq:dyn2D} \\
    r(\vx,t) &= & \frac{U(\vx,t)^2}{1+k\rho 
    \int^\infty_{-\infty}\int^{\infty}_{-\infty} d\vx' U(\vx',t)^2},
\label{eq5.2}
\end{eqnarray}
where $\tau$ is the time constant and
$\rho$ the neural density, and the recurrent interactions are chosen to be
\begin{equation}
    J(\vx,\vx')=\frac{A}{2\pi a^2}
    \exp\left[-\frac{(\vx-\vx')^2}{2a^2}\right],
\label{eq5.3}
\end{equation}
where $(\vx-\vx')^2 = (x_1-x_1')^2+(x_2-x_2')^2$ is the Euclidean distance
between $\vx$ and $\vx'$.

It can be checked that when $I_{\rm ext}(\vx,t)=0$ and
$0<k<k_c\equiv A^2\rho/(32\pi a^2)$, the network holds a continuous
family of stationary states given by
\begin{eqnarray}
    \tilde{U}(\vx|\vz) & = & U_0\exp\left[-\frac{(\vx-\vz)^2}{4a^2}\right],
\label{eq5.4} \\
    \tilde{r}(\vx|\vz) & = & r_0\exp\left[-\frac{(\vx-\vz)^2}{2a^2}\right],
\label{eq5.5}
\end{eqnarray}
where $U_0=[1+(1-k/k_c)^{1/2}]A/(8\pi a^2k)$ and
$r_0=[1+(1-k/k_c)^{1/2}]/(4\pi a^2 k\rho)$. The parameter
$\vz=(z_1,z_2)$ is a free variable, indicating the location of the
bump. We note the differences in $J(\vx,\vx')$, $k_c$, $U_0$ and $r_0$
when compared with their values in the 1D case.

We analyze the stability of the bump states, 
and consider the network state $U(\vx,t)=\tilde{U}(\vx|\vz)+\delta U(\vx,t)$.
After linearizing Eq.~(\ref{eq:dyn2D}) at $\tilde{U}(\vx|\vz)$, we get
\begin{equation}
    \tau\frac{\partial}{\partial t}\delta U(\vx,t)
    =\int^\infty_{-\infty} \int^\infty_{-\infty}
    d\vx'F(\vx,\vx'|\vz)\delta U(\vx',t)-\delta U(\vx,t),
\label{eq5.6}
\end{equation}
where the interaction kernel is given by
\begin{eqnarray}
    F(\vx,\vx'|\vz) & = &
    \frac{2}{a^2\pi}\exp\left[-\frac{(\vx-\vx')^2}{2a^2}\right]
    \exp\left[-\frac{(\vx'-\vz)^2}{4a^2}\right],
    \nonumber \\
    & - & {\frac {1+\sqrt{1-k/k_c}} {2\pi a^2}}\exp
    \left[-\frac{(\vx-\vz)^2}{4a^2}\right]
    \exp\left[-\frac{(\vx'-\vz)^2}{4a^2}\right].
\label{eq5.8}
\end{eqnarray}
We can construct the right eigenfunction of $F(\vx,\vx'|\vz)$
by using the product of
the eigenfunctions $u_n(x|z)$ in the 1D case, i.e.,
\begin{equation}
    u_{m,n}(\vx|\vz)=u_m(x_1|z_1)u_n(x_2|z_2), ~~m,n=0,1,2,\ldots.
\end{equation}
The corresponding eigenvalues are given by (see Appendix F)
\begin{eqnarray}
    \lambda_{0,0} & = & \lambda_0 \\
    \lambda_{m,0} & =& \lambda_m, ~~ m\neq 0, \\
    \lambda_{0,n} & = & \lambda_n, ~~ n\neq 0, \\
    \lambda_{m,n}& = & \lambda_m \lambda_n, ~~m\neq 0,n\neq 0,
\end{eqnarray}
where $\lambda_m$ are the eigenvalues in the 1D case
given by Eqs.~(\ref{eq:l0}) and (\ref{eq:ln}).

The eigenfunction $u_{0,0}(\vx|\vz)$ corresponds to the
amplitude mode of the 2D bump state of the network.
$u_{1,0}(\vx|\vz)$ corresponds to the displacement of the bump position
along the coordinate $z_1$
and $u_{0,1}(\vx|\vz)$ the positional shift along $z_2$.
A linear combination of $u_{1,0}(\vx|\vz)$ and $u_{0,1}(\vx|\vz)$, i.e.,
$c_1 u_{1,0}(\vx|\vz)+c_2 u_{0,1}(\vx,\vz)$,
with $c_1$ and $c_2$ being constants and satisfying  $c_1^2+c_2^2=1$,
corresponds to the positional shift of the bump 
along the $\vc=(c_1,c_2)$ direction. 
Interestingly, the eigenvalue for $c_1 u_{1,0}(\vx|\vz)+c_2 u_{0,1}(\vx,\vz)$
is 1, indicating that the bump is neutrally stable
in the two-dimensional space $\vz$.

Analogous to the 1D case, we can develop a perturbation method to
solve the network dynamics, e.g.,
by using the 2D wave functions, $v_{m,n}(\vx|\vz)=v_n(x_1|z_1)v_n(x_2|z_2)$,
of the quantum harmonic oscillators as the basis functions.
A more comprehensive treatment of the problem
involves transforming the basis functions to polar coordinates; 
it will be presented elsewhere.
Here, as an illustration, we will
present only the result for the weak input limit
in the rectilinear coordinates,
which corrsponds to projecting the network dynamics
on a single displacement mode.
We choose the external input to be
\begin{equation}
    I_{\rm ext}(\vx,t) =
    \alpha U_0\exp\left[-{\frac {(\vx-\vz^0)^2} {4a^2}}\right]
    +\sigma \eta(\vx,t),
\label{eqD1}
\end{equation}
where $\vz^0$ represents the stimulus location.

Suppose the bump position is $\vz(t)$ at time $t$.
Then, the stimulus will pull the bump towards the location $\vz^0$.
In the limit of weak inputs, the network dynamics is dominated
by the positional shift of the bump along the straight line through $\vz^0$. 
As derived in Appendix F,
\begin{equation}
    \frac{d(\vz-\vz^0)}{dt}=\frac{\alpha}{\tau}(\vz-\vz^0)\exp\left[
    -\frac{(\vz-\vz^0)^2}{8a^2}\right]+\sqrt{2D}\eta(t),
\label{eq:velocity2D}
\end{equation}
where $D=\sigma^2/(U_0^2\tau^2\pi)$.
This equation describes how the distance between the bump and the stimulus
varies with time. It has the same form as Eq.~(\ref{OU})
except that $z-z^0$ is replaced by $\vz-\vz^0$.
Thus, the network exhibits the same dynamical properties
as observed in the 1D case.
Fig.~10 compares the theoretical predictions
and the simulation results of the reaction time of the 2D CANN.
As expected, they agree with each other very well with small jump sizes.

\vskip 0.5cm
\begin{figure*}[htb]
\begin{center}
\epsfig{file=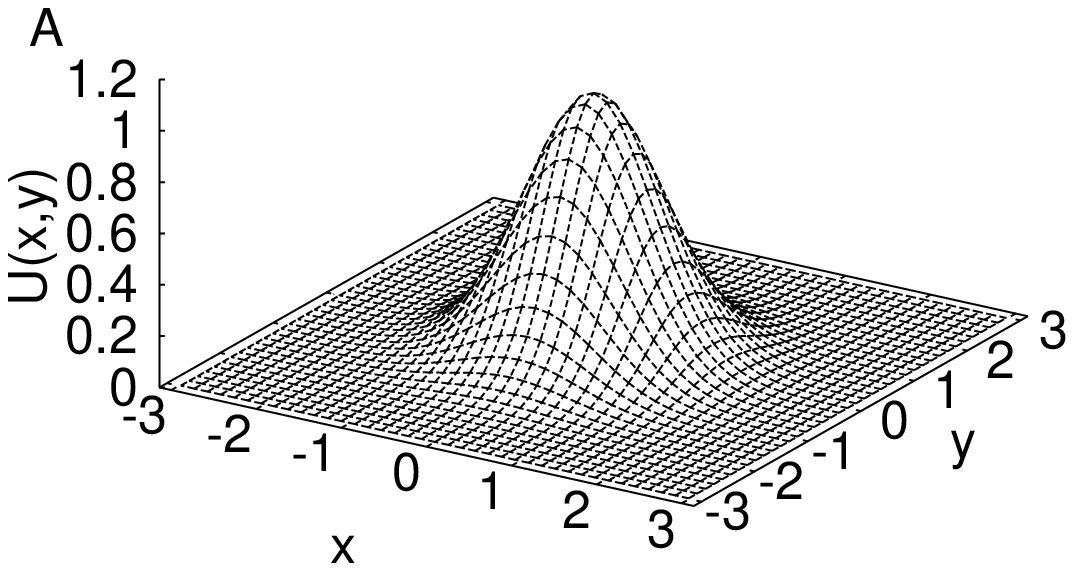,width=6cm}
\\
\vskip 1cm
\epsfig{file=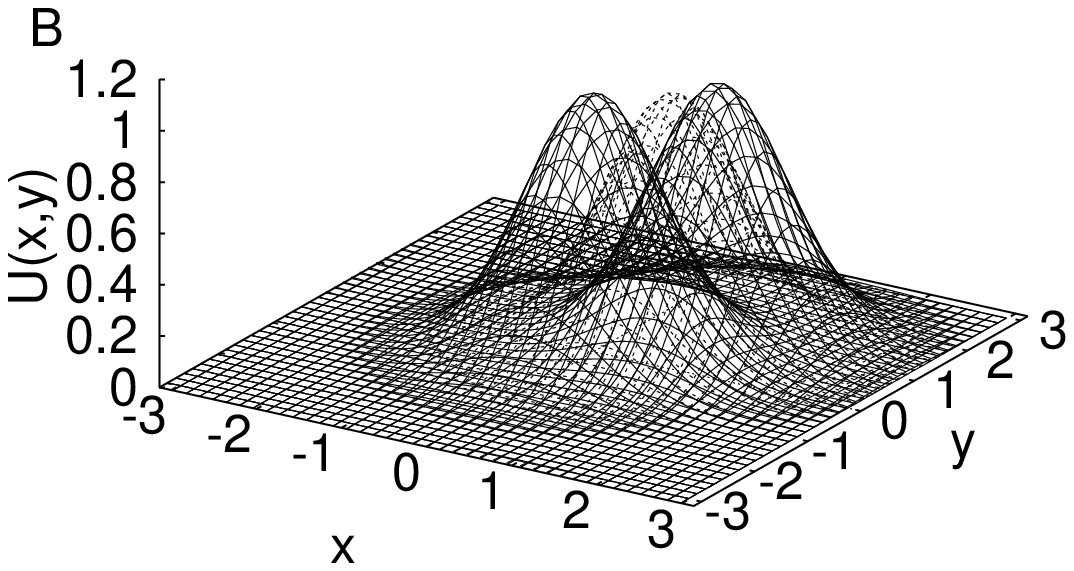,width=6cm}
\\
\vskip 1cm
\epsfig{file=fig10c.eps,width=6cm}
\caption{(A) The stationary state of the 2D CANN.
(B) Illustrating the tracking process
of the network; (C)
The reaction time vs. the jump size.
The simulation result is compared with the theoretical prediction. 
Parameters: $N = 40\times 40 $, $k = 0.5$, $a = 0.5$, $\tau = 1$,
$A = \sqrt{2}\pi a^2$, $\rho = N/(2\pi)^2$, $\alpha = 0.05$ 
and $\theta = \sqrt{2}\pi/N$.}
\end{center}
\end{figure*}

\section{Discussion and Conclusions}

The idea that the landscape of the energy function of a CANN
has a canyon structure,
which facilitates the tracking capacity of a neural system
has long been recognized (see, e.g., 
\cite{Amari77,Hansel1998,Brody03,Wu05}).
However, a detailed and rigorous analysis of this property has been lacking.
The main challenge here 
is to solve the network dynamics analytically, elucidating
how the shape of the network state is affected during the tracking.
In this study, by using a simple analytically solvable model,
we systematically investigate this issue.
There are two parts to our main results. 
First, we show that the dynamics of a CANN can be decomposed into different
motion modes, corresponding to distortions in 
the height, position, width, skewness and other higher-order distortions
in the shape of the bump states. 
We demonstrate that with a CANN, it is the positional shift
that dominates the network dynamics,
with other motion modes having high-order contributions.
Second, we develop a time-dependent
perturbation approach to approximate the network dynamics.
Geometrically, this corresponds to projecting the network dynamics
onto its dominating motion modes.
Simulation results confirm that our method works very well.
Two interesting phenomena in the tracking performance of a CANN are observed,
namely, the maximum trackable speed
and the logarithmic reaction time of the network.
Since both properties are associated with the unique dynamics of a CANN,
they may be tested in practice and serve as clues
for checking the application of a CANN in neural systems.

To facilitate an analytical solution
and hence to describe the network dynamics clearly,
we have used a simple firing-rate-based model
with the inhibition effect in the form of divisive normalization,
and the generality of our results may thus be a concern.
In particular, we should consider 
whether the mechanism of divisive normalization is plausible,
whether tracking behavior can be observed
in models with different inhibitory mechanisms,
and how the perturbation method developed in this paper
can be used in other types of networks.

First, we argue that a neural system can have resources
to implement divisive normalization~\cite{Grossberg88,Heeger93,Deneve99}.
Let us consider a network in which all excitatory neurons
are connected to a pool of inhibitory neurons.
If the time constant of inhibitory neurons is much smaller
than that of excitatory neurons, and if 
these inhibitory neurons inhibit the activity of the excitatory neurons 
in a uniform shunting way,
then the effect of divisive normalization can be achieved.

Second, even if there exists uncertainty of the model,
the main conclusions of tracking behavior in this work
are applicable to general cases.
This is because our calculation is based on the fact that
the dynamics of a CANN is largely dominated by the positional shift
of the network states,
a property coming from the translational invariance
of the neuronal interactions, 
rather than from the inhibition mechanism.
To see this point clearly, let us consider the dynamics of a network
having the same form as Eq.~(\ref{eq:dyn}),
but the exact form of the inihibition mechansim remains unspecified.
We consider $J(x-x')$ and $r(x)$ that are properly chosen
(e.g., $J(x-x')$ is of the Mexcian-hat shape
and $r(x)$ a sigmoid function of the inputs),
so that the network holds a continuous family of stationary states,
denoted as $\tilde{U}(x|z)$ and $\tilde{r}(x|z)$, 
with $z$ representing the bump position. This implies that 
\begin{equation}
    \tilde{U}(x|z)=\rho\int^\infty_{-\infty}
    dx' J(x-x')\tilde{r}(x'|z).
\end{equation}
By differentiating both sides of the above equation with respect to $z$,
we get
\begin{eqnarray}
    \frac{\partial\tilde{U}(x|z)}{\partial z} & = &
    \rho\int^\infty_{-\infty} dx'\int^\infty_{-\infty} dx''
    J(x-x')\frac{\partial \tilde{r}(x'|z)}
    {\partial \tilde{U}(x''|z)}
    \frac{\partial\tilde{U}(x''|z)}{\partial z}
    \nonumber \\
    &=& \int^\infty_{-\infty} dx''
    F(x,x''|z)\frac{\partial\tilde{U}(x''|z)}{\partial z}.
\end{eqnarray}
Thus, the mode of the positional shift of the bump,
$\partial\tilde{U}(x|z)/\partial z$,
is the eigenfunction of the interaction kernel $F$ 
with the eigenvalue equal to 1.
We should observe similar tracking behavior of the network,
once $J(x-x')$ and $r(x)$ are properly chosen
to ensure that all the other eigenvalues are less than one.

Third, the perturbation method developed in this paper
can be used to analyze other types of networks with tracking behavior.
Consider appropriate choices of $J(x-x')$ and $r(x)$
that can hold a continuous family of bump states
centered at position $z$
(for example, $J(x-x')$ having a Mexican-hat shape
and $r(U)$ a sigmoid function).
Then, the dynamics of the bump
can be described by Eq.~(\ref{eq:fluc}), with
\begin{equation}
    F(x,x'|z)=\int^\infty_{-\infty}dx'J(x-x')
    \frac{\partial r(\tilde U(x'|z))}{\partial\tilde U(x'|z)}.
\end{equation}
Using the basis functions of the quantum harmonic oscillator,
we can obtain the matrix elements $F_{mn}$
from Eq.~(\ref{eq:fmn_def}).
While elegant expressions such as those in Eq.~(\ref{eq:fmn})
may not be available in the general case,
we can nevertheless compute $F_{mn}$ numerically.
Perturbation dynamics similar to
Eqs.~(\ref{eq:perturb}) and (\ref{eq:velocity})
can be worked out following Appendix D.
Indeed, we have considered networks with Mexican-hat-shaped interactions.
Tracking dynamics with a logarithmic reaction time 
and a maximum trackable speed is observed,
and the convergence of the perturbation method
is equally remarkable.
These results will be reported elsewhere.

The tracking dynamics of a CANN has also been studied by other authors.
In particular, Zhang proposed a
mechanism of using asymmetrical recurrent interactions to drive the bump, 
so that
the shape distortion would be minimized \cite{Zhang96}.
Xie {\it et al} further proposed a double ring network model
that uses the inputs from 
the vestibular system in mammals to achieve these asymmetrical
interactions~\cite{Xie02}.
This mechansim works well for the head-direction system, but
it is unclear if it can be achieved in other neural systems.
For instance, in the visual and hippocampal systems,
it is often assumed that the bump movement is directly driven 
by external inputs (see, e.g., \cite{Samsonovich97,Berry99,Fu01}),
and the distortion of the bump is inevitable
(indeed the bump distortions in \cite{Berry99,Fu01} 
are associated with visual perception).
The contribution of this study is
that we quantify how the distortion of the bump's shape
affects the network tracking performance, and we obtain
the maximum trackable speed of the network.
We hope that this study will enrich our knowledge
about the tracking dynamics of a CANN, 
which may differ in different brain functions.
Apart from the tracking capacity,
recent experimental observations have suggested that
CANNs may have other important roles in neural cognitive functions. 
For example, the highly structured state space of
a CANN may provide a neural basis 
of encoding the categorization relationships 
of objects~\cite{Jastorff06,Graf06}.
It is quite possible that the distance between
two memory states in the canyon defines the perceptual similarity
between the two objects. 
We will explore these issues in the future,
and  we expect that the theoretical model
and approaches developed in this work
will provide valuable tools.

\section*{Acknowledgment}
This work was partially supported
by the Research Grants Council of Hong Kong
(Grant Nos. HKUST 603606, 603607 and 604008), and BBSRC (BB/E017436/1)
and the Royal Society of UK.

\section*{Appendix A: The Hermite Polynomials}

Since the Hermite polynomials are frequently used
in the calculations of this paper,
here we summarize their main properties for the convenience of the reader.

The $n$th-order Hermite polynomial is defined as
\begin{equation}
    H_n(x)=(-1)^n\exp(x^2)\left(\frac{d}{dx}\right)^n\exp(-x^2).
\label{eq:hermite}
\end{equation}
The first five Hermite polynomials are given by
$H_0(x)=1$, $H_1(x)=2x$, $H_2(x)=4x^2-2$, $H_3(x)=8x^3-12x$,
$H_4(x)=16x^4-48x^2+12$.
For the analysis in this paper,
it is instructive to note that
the Hermite polynomials satisfy the following properties.
\begin{itemize}
\item
Recursion relation:
\begin{eqnarray}
    H_{n+1}(x) &= & 2xH_n(x)-H'_n(x),
\label{eq:hn+1} \\
    H'_n(x) & = & 2nH_{n-1}(x),
\label{eq:dhn} \\
    H_n(x+y) & = & \sum_{k=0}^{n}
    \left(\begin{array}{c}n\\k\end{array}\right)H_k(x)(2y)^{n-k}.
\label{eq:hx+y}
\end{eqnarray}
\item
Contour integral representation:
\begin{equation}
    H_n(x)=n!\oint \frac{dt}{2\pi it^{n+1}}e^{2tx-t^2},
\label{eq:contour}
\end{equation}
with the contour encircling the origin.
\end{itemize}

\section*{Appendix B: The Interaction Kernel $F(x,x'|z)$}

Linearizing Eq.~(\ref{eq:dyn}) at $\tilde U(x|z)$, we arrive at
\begin{eqnarray}
    \tau\frac{\partial}{\partial t}\delta U(x)
    &=&\rho\int^\infty_{-\infty}dx'J(x,x')
    \left[\frac{2U(x')}{B}\delta U(x')\right.
    \nonumber\\
    &-&\left.\frac{U(x')^2}{B^2}\int^\infty_{-\infty} dx''
    2k\rho U(x'')\delta U(x'')\right]-\delta U(x),
\end{eqnarray}
where we have dropped the dependence on $t$ for brevity, and
$B=1+k\rho\int^{\infty}_{-\infty} dx'\tilde U(x')^2=\rho AU_0/\sqrt
2$. Interchanging the dummy variables $x'$ and $x''$ in the last
term, we obtain Eqs.~(\ref{eq:fluc}) and (\ref{eq:kernel}), with
\begin{eqnarray}
    F(x,x'|z) & = & {\frac {2} {a\sqrt{\pi}}}
    \exp \left[-\frac{(x-x')^2}{2a^2}\right]
    \exp \left[-\frac{(x'-z)^2}{4a^2}\right]
    \nonumber \\
    & - & {\frac {1+\sqrt{1-k/k_c}} {\sqrt{2\pi}a}}
    \exp \left[-\frac{(x-z)^2}{4a^2}\right]
    \exp\left[-\frac{(x'-z)^2}{4a^2}\right].
\end{eqnarray}
By substituting the expressions of $U(x)$, $J(x,x')$ and $r(x)$,
we get Eq.~(\ref{eq:kernel}).

To calculate the matrix elements $F_{mn}$ in Eq.~(\ref{eq:fmn_def}),
we denote 
\begin{equation}
    g_n(x|z)=\int^\infty_{-\infty} dx' F(x,x'|z)v_n(x'|z).
\end{equation}
From Eqs.~(\ref{eq:kernel}) and (\ref{eq:basis}),
\begin{eqnarray}
    g_n(x|z)&=&\frac{(-1)^n2(\sqrt{2}a)^{n-\frac{1}{2}}}
    {a\sqrt{\pi^{3/2}n!2^n}}\int^\infty_{-\infty} dx'
    \exp\left[-\frac{(x-x')^2}{2a^2}\right]
    \left(\frac{d}{dx'}\right)^n\exp\left[-\frac{(x'-z)^2}{2a^2}\right]
    \nonumber \\
    & &-\frac{(1+\sqrt{1-k/k_c})(-1)^n(\sqrt{2}a)^{n-\frac{1}{2}}}
    {a\sqrt{\pi^{3/2}n!2^{n+1}}}
    \exp \left[-\frac{(x-z)^2}{4a^2}\right] \nonumber \\
    & & \times
    \int^\infty_{-\infty} dx'\left(\frac{d}{dx'}\right)^n
    \exp\left[-\frac{(x'-z)^2}{2a^2}\right].
\label{eqA2}
\end{eqnarray}
The second term vanishes for $n\ge 1$.
Hence, by evaluating the two integrals explicitly for $n=0$, we have
\begin{equation}
    g_0(x|z)=(1-\sqrt{1-k/k_c})v_0(x|z).
\label{eqA3}
\end{equation}
Making use of the orthonormality of the basis $v_n(x|z)$,
we obtain the first line of Eq.~(\ref{eq:fmn}).

Next, we consider the case $n\geq 1$,
focusing on the first term.

Utilizing the identity $(-d/dz)^ne^{-(x-z)^2/2a^2}
=(d/dx)^ne^{-(x-z)^2/2a^2}$, we have 
\begin{equation}
    g_n(x|z)
    =\frac{(-1)^n2(\sqrt{2}a)^{n-\frac{1}{2}}}{a\sqrt{\pi^{3/2}n!2^n}}
    \left(-\frac{d}{dz}\right)^n\int^\infty_{-\infty} dx'
    \exp\left[-\frac{(x-x')^2}{2a^2}\right]
    \exp\left[-\frac{(x'-z)^2}{2a^2}\right].
\label{eq:gnx}
\end{equation}
Evaluating the integral explicitly, we find that 
\begin{equation}
    g_n(x|z)=\frac{(-1)^n2(\sqrt{2}a)^{n-\frac{1}{2}}}
    {\sqrt{\pi^{1/2}n!2^n}}\left(\frac{d}{dx}\right)^n
    \exp\left[-\frac{(x-z)^2}{4a^2}\right].
\label{eqA6}
\end{equation}
We are now ready to derive $F_{mn}$ from $g_n(x|z)$: 
\begin{equation}
    F_{mn}=\int^\infty_{-\infty} dx v_m(x|z)g_n(x|z).
\end{equation}
From Eqs.~(\ref{eq:basis}) and (\ref{eq:gnx}),
\begin{eqnarray}
    F_{mn}& = &
    2\frac{(-1)^m(\sqrt{2}a)^{m-\frac{1}{2}}}{\sqrt{\pi^{1/2}m!2^m}}
    \frac{(-1)^n(\sqrt{2}a)^{n-\frac{1}{2}}}{\sqrt{\pi^{1/2}n!2^n}}
    \nonumber \\
    &&\times \int^\infty_{-\infty} dx
    \exp\left[\frac{(x-z)^2}{4a^2}\right]
    \left(\frac{d}{dx}\right)^m
    \exp\left[-\frac{(x-z)^2}{2a^2}\right]
    \left(\frac{d}{dx}\right)^n\exp\left[-\frac{(x-z)^2}{4a^2}\right].
\end{eqnarray}
Using the definition of the Hermite polynomials in Eq.~(\ref{eq:hermite}),
this simplfies to
\begin{equation}
    F_{mn}=2\frac{(-1)^m(\sqrt{2}a)^{m+n-1}}{\sqrt{\pi m!n!2^{m+n}}}
    \frac{1}{(2a)^n}\int^\infty_{-\infty} dx
    H_n\left(\frac{x-z}{2a}\right)
    \left(\frac{d}{dx}\right)^m\exp\left[-\frac{(x-z)^2}{2a^2}\right].
\end{equation}
Applying integration by parts $m$ times yields 
\begin{equation}
    F_{mn}=2\frac{(\sqrt{2}a)^{m+n-1}}{\sqrt{\pi m!n!2^{m+n}}}
    \frac{1}{(2a)^n}\int^\infty_{-\infty} dx
    \exp\left[-\frac{(x-z)^2}{2a^2}\right]
    \left(\frac{d}{dx}\right)^m H_n\left(\frac{x-z}{2a}\right).
\label{eqA7}
\end{equation}
Applying Eq.~(\ref{eq:dhn}) $m$ times yields, for $n\geq m$,
\begin{equation}
    F_{mn}
    = 2\frac{(\sqrt{2}a)^{m+n-1}}{\sqrt{\pi m!n!2^{m+n}}} \frac{1}{(2a)^n}
    \frac{n!}{a^m(n-m)!}
    \int^\infty_{-\infty} dx \exp\left[-\frac{(x-z)^2}{2a^2}\right]
    H_{n-m}\left(\frac{x-z}{2a}\right),
\label{eqA8}
\end{equation}
and $F_{mn}=0$ otherwise.

Introducing the contour integral representation in Eq.~(\ref{eq:contour}) 
leads to 
\begin{equation}
    F_{mn}=\frac{\sqrt{2}\sqrt{n!}}{2^na\sqrt{\pi m!}}
    \int^\infty_{-\infty} dx \exp\left[-\frac{(x-z)^2}{2a^2}\right]
    \oint \frac{dt}{2\pi it^{n-m+1}} \exp\left[-t^2
    +t\left(\frac{x-z}{a}\right)\right].
\end{equation}
Interchanging the order of integrating $x$ and $t$, we have 
\begin{equation}
    F_{mn}=2^{1-n} \sqrt{\frac{n!}{m!}}
    \oint \frac{dt}{2\pi it^{n-m+1}} \exp\left(-\frac{t^2}{2}\right).
\end{equation}
According to the residue theoreom in complex analysis,
the contour integral can be reduced to the coefficient of $t^{n-m}$
in the Laurent expansion of $\exp(-t^2/2)$,
which is equal to $(-1/2)^{(n-m)/2}/[(n-m)/2]!$
when $n-m$ is even and 0 otherwise.
Hence, we have derived $F_{mn}$
in the second line of Eq.~(\ref{eq:fmn}).

Other than the above cases,
all other elements $F_{mn}$ are zero.
Hence we have derived Eq.~(\ref{eq:fmn}) completely.

As an illustration, we write down
the elements of ${\mathbf F}$ in the first four rows and columns,
\begin{equation}
    {\mathbf F}=\pmatrix{
    1-\sqrt{1-k/k_c} & 0 & -{\frac {\sqrt{2}} {4}} & 0 & \ldots \cr
    0 & 1 & 0  &  -{\frac {\sqrt{6}} {8}} & \ldots \cr
    0 & 0 & {\frac 1 2} & 0 & \ldots \cr
    0 & 0 & 0 & {\frac 1 4} & \ldots \cr
    \ldots & \ldots & \ldots & \ldots & \ldots \cr}.
\label{eqA11}
\end{equation}
Since all elements of the matrix ${\mathbf F}$
in the lower half of the diagonal line vanish,
the eigenvalues of ${\mathbf F}$ are given by
the diagonal elements, namely, Eqs.~(\ref{eq:l0}) and (\ref{eq:ln}).
It is straightforward to check that
the first four right eigenvectors of $F_{mn}$ are given by
\begin{eqnarray}
    \vq_0 & = & (1,0,0,\ldots)^T,
\label{eqA14}\\
    \vq_1 & = & (0,1,0,\ldots)^T,
\label{eqA15} \\
    \vq_2 & = & ({\frac {\sqrt{1/2}} {D_0}},0,
    \frac{1-2\sqrt{1-k/k_c}}{D_0},0,\ldots)^T,
\label{eqA16}\\
\vq_3 & = & (0,\sqrt{\frac{1}{7}},0,\sqrt{\frac{6}{7}},0,\ldots)^T,
\label{eqA17}
\end{eqnarray}
where $D_0$ is given in Eq.~(\ref{eq:d0})
and all the omitted elements are zero.
The right eigenfunctions of $F(x,x'|z)$,
with eigenvalues $\lambda_n$,
are given by
\begin{equation}
    u^R_n(x|z)=\sum_l q_{nl}v_l(x|z).
\end{equation}
This can be verified by considering the inner product
\begin{equation}
    \int^\infty_{-\infty} dx' F(x,x'|z)u^R_k(x'|z)
    =\int^\infty_{-\infty} dx'
    \sum_{mn}v_m(x|z)F_{mn}v_{n}(x'|z)\sum_l q_{kl} v_l(x'|z).
\end{equation}
Making use of the orthonormality of $v_l(x'|z)$, we have 
\begin{equation}
    \int^\infty_{-\infty} dx' F(x,x'|z)u^R_k(x'|z)
    =\sum_{ml}v_m(x|z)F_{ml}q_{kl}.
\end{equation}
Since $\vq_k$ is an eigenvector of ${\mathbf F}$
with eigenvalue $\lambda_k$,
\begin{equation}
    \int^\infty_{-\infty} dx' F(x,x'|z)u^R_k(x'|z)
    =\lambda_k \sum_m v_m(x|z)q_{km}
    =\lambda_k u^R_k(x|z).
\label{eqA18}
\end{equation}
The first four right eigenfunctions of ${\mathbf F}$
are presented in Eqs.~(\ref{eq:u0})-(\ref{eq:u3}).

\section*{Appendix C: The Network Dynamics in the Weak Input Limit}

We let $U(x,t)=\tilde{U}(x|z(t))$
and multiply both sides of Eq.~(\ref{eq:dyn}) by $v_1(x|z(t))$,
and then we integrate over $x$. Since
\begin{equation}
    \frac{\partial\tilde U(x|z(t))}{\partial t}
    =\frac{U_0\sqrt{(2\pi)^{1/2}a}}{2a}\frac{dz}{dt}
    v_1(x|z(t)),
\label{eqB1}
\end{equation}
the left-hand side of the resultant equation becomes
\begin{equation}
    L.H.S.=\frac{\tau U_0\sqrt{(2\pi)^{1/2}a}}{2a}\frac{dz}{dt}.
\end{equation}
On the right hand side of Eq.(\ref{eq:dyn}),
the first term arises from the external input,
consisting of both signal and noise components
according to Eq.~(\ref{eq:input}).
The signal component,
when projected onto the component $v_1(x|z(t))$, leads to
\begin{equation}
    R.H.S._{\rm signal}
    =\frac{\alpha U_0}{\sqrt{(2\pi)^{1/2}a}}
    \int^\infty_{-\infty} dx\exp\left[-\frac{(x-z_0)^2}{4a^2}\right]
    \left(\frac{x-z}{a}\right)\exp\left[-\frac{(x-z)^2}{4a^2}\right].
\label{eqB2}
\end{equation}
Explicit integration yields
\begin{equation}
    R.H.S._{\rm signal}
    =\alpha U_0\sqrt{(2\pi)^{1/2}a}\frac{z_0-z}{2a}
    \exp\left[-\frac{(z_0 - z)^2}{8a^2}\right].
\end{equation}
The noise component leads to
\begin{equation}
    R.H.S._{\rm noise}=\sigma\eta_1(t),
\end{equation}
where $\eta_1$ is given in Eq.~(\ref{eq:noise1})
and satisfies Eq.~(\ref{eq:noiseav}).

The second and third terms of Eq.~(\ref{eq:dyn}) vanish
owing to the fact that $\tilde U(x|z(t))$
is the stationary state solution of Eq.~(\ref{eq:dyn})
in the weak input limit,
independent of the bump position $z(t)$.

Combining the above results, we get Eq.~(\ref{OU}).

\section*{Appendix D: The Perturbation Method}

\subsection*{D.1 The Perturbation Equation}

Using Eq.~(\ref{eq:basis}), we have
\begin{equation}
    \frac{d}{dt}v_n(x|z)=\frac{1}{\sqrt{(2\pi)^{1/2}an!2^n}}
    \exp\left[-\frac{(x-z)^2}{4a^2}\right]
    \left[\frac{x-z}{2a^2}H_n\left(\frac{x-z}{\sqrt{2}a}\right)
    -\frac{1}{\sqrt{2}a}H'_n\left(\frac{x-z}{\sqrt{2}a}\right)\right]
    \frac{dz}{dt}.
\label{eqC3}
\end{equation}
By virtue of Eq.~(\ref{eq:hn+1}) and (\ref{eq:dhn}),
\begin{equation}
    \frac{d}{dt}v_n(x|z)
    =\left[\sqrt{n+1}v_{n+1}(x|z)-\sqrt{n}v_{n-1}(x|z)\right]
    \frac{1}{2a}\frac{dz}{dt}.
\label{eqC4}
\end{equation}
Similarly, since $\tilde{U}(x|z)=U_0\sqrt{(2\pi)^{1/2}a}v_0(x|z)$,
\begin{equation}
    \frac{\partial}{\partial t}\tilde{U}(x|z)
    =U_0\sqrt{(2\pi)^{1/2}a}v_1(x|z)\frac{1}{2a}\frac{dz}{dt}.
\label{eqC5}
\end{equation}
The neuronal interaction and relaxation terms
cancel each other at the steady state.
Hence, in determining the dynamics, only the linearized terms
need to be considered,
\begin{equation}
    \rho\int^\infty_{-\infty} dx' J(x,x')r(x'|z)-U(x|z)
    \approx \sum_n a_n \sum_m v_m(x|z)F_{mn}-\sum_n a_n v_n(x|z).
\label{eqC6}
\end{equation}
Substituting Eqs.~(\ref{eqC4}-\ref{eqC6}) into Eq.~(\ref{eq:dyn}),
we obtain
\begin{eqnarray}
    && \tau \sum_n \frac{da_n}{dt}v_n
    +\frac{\tau}{2a}\left\{U_0\sqrt{(2\pi)^{1/2}a}v_1
    +\sum_n a_n \left[\sqrt{n+1}v_{n+1}-\sqrt{n}v_{n-1}\right]\right\}
    \frac{dz}{dt} \nonumber \\
    &=&\sum_n a_n \sum_m v_m F_{mn}-\sum_n a_n v_n +\sum_n I_n v_n.
\label{eqC7}
\end{eqnarray}

Making use of the orthonormality and completeness of $v_n(x|z)$,
the coefficients in Eq.~(\ref{eqC7}) can be equated term by term.
Hence, we get
\begin{equation}
    \tau \frac{da_n}{dt}
    +\frac{\tau}{2a}\left[ U_0\sqrt{(2\pi)^{1/2}a}\delta_{n1}
    +\sqrt{n}a_{n-1}-\sqrt{n+1}a_{n+1}\right]
    \frac{dz}{dt}=\sum_k F_{nk}a_k-a_n+I_n.
\label{eqC8}
\end{equation}
By further using Eq.~(\ref{eq:fmn}), the above equation
is re-organized as Eq.~(\ref{eq:perturb}).

\subsection*{D.2 The Velocity Equation}

The peak position $z(t)$ of the bump
is determined from the self-consistent condition in Eq.~(\ref{eq:mass}).
Without loss of generality, we consider $z=0$ 
and substitute the expression of $v_n(x|z)$ into Eq.~(\ref{eq:mass}),
\begin{equation}
    \sum_{n=1,\rm odd}\frac{a_n}{\sqrt{n!2^n}}
    \int^\infty_{-\infty} dx \exp\left(-\frac{x^2}{4a^2}\right)
    xH_n\left(\frac{x}{\sqrt{2}a}\right)=0.
\end{equation}
Integrating by parts and using the identity in Eq.~(\ref{eq:dhn}), we have 
\begin{equation}
    \sum_{n=1,\rm odd}\frac{na_n}{\sqrt{n!2^n}}
    \int^\infty_{-\infty} dx \exp\left(-\frac{x^2}{4a^2}\right)
    H_{n-1}\left(\frac{x}{\sqrt{2}a}\right)=0.
\end{equation}
The integral can be performed
by using the contour integral representation
of the Hermite polynomial in Eq.~(\ref{eq:contour}),
\begin{equation}
    \sum_{n=1,\rm odd}\frac{n!a_n}{\sqrt{n!2^n}}
    \int^\infty_{-\infty} dx \oint \frac{dt}{2\pi i t^n}
    \exp\left(-t^2-\frac{x^2}{4a^2}
    +\frac{2tx}{\sqrt{2}a}\right)=0.
\end{equation}
Interchanging the order of integration yields 
\begin{equation}
    \sum_{n=1,\rm odd}\frac{n!a_n}{\sqrt{n!2^n}}
    \oint \frac{dt}{2\pi i t^n}\exp(t^2)=0.
\end{equation}
The contour integral is equal to the coefficient of $t^{n-1}$ in the series
expansion of $\exp(t^2)$, which is $[(n-1)/2]!$. Hence, 
\begin{equation}
    \sum_{n=1,\rm odd}\sqrt{\frac{n!!}{(n-1)!!}}a_n=0.
\label{eq:cm}
\end{equation}
Multiplying both sides of Eq.~(\ref{eq:perturb})
by $\sqrt{n!!/(n-1)!!}$, and summing
over odd $n$, the time derivatives of $a_n$ disappear, leaving behind
\begin{eqnarray}
    &&0=\frac{1}{\tau}\sum_{n=1,\rm odd}
    \sqrt{\frac{n!!}{(n-1)!!}}I_n
    \nonumber \\
    && -\left[U_0\sqrt{(2\pi)^{1/2}a}
    +\sum_{n=1,\rm odd}\sqrt{\frac{n!!}{(n-1)!!}}
    \left(\sqrt{n}a_{n-1}-\sqrt{n+1}a_{n+1}\right)\right]
    \frac{1}{2a}\frac{dz}{dt} \nonumber \\
    && +\frac{1}{\tau}\sum_{n=1,odd}\sqrt{\frac{n!!}{(n-1)!!}}
    \sum_{r=0}^{\infty}\sqrt{\frac{(n+2r)!}{n!}}
    \frac{(-1)^r}{2^{n+3r-1}r!}a_{n+2r}.
\end{eqnarray}
Collecting terms in the summations, we have 
\begin{eqnarray}
    &&\!\!\!\!\left\{U_0\sqrt{(2\pi)^{1/2}a}+
    \sum_{n=0,\rm even}\left[\sqrt{\frac{(n+1)!!}{(n)!!}}\sqrt{n+1}
    -\sqrt{\frac{(n-1)!!}{(n-2)!!}}\sqrt{n}\right]a_n\right\}
    \frac{1}{2a}\frac{dz}{dt} \nonumber \\
    &=&\!\!\!\!\frac{1}{\tau}\sum_{n=1,\rm odd}\!\!
    \sqrt{\frac{n!!}{(n-1)!!}}(I_n\!-\!a_n)
    \!+\!\frac{1}{\tau}\sum_{m=1,\rm odd}\!\!
    \sqrt{m!}\left[\sum_{r=0}^{(m-1)/2}\!\!
    \frac{(-1)^r}{2^{m+r-1}(m-2r-1)!!r!}\right]a_m.
\end{eqnarray}
Note that for odd $m$, $(m-2r-1)!!=2^{(m-2r-1)/2}[(m-1)/2-r]!$,
so that the summation over $r$ in the last term is proportional to
the binomial expansion of $[1+(-1)]^{(m-1)/2}$, which is equal to
$1$ for $m=1$, and vanishes otherwise.
Therefore, Eq.~(\ref{eq:velocity}) is obtained.

\section*{Appendix E: The Components of the External Input}

We first calculate the component of the signal term in Eq.~(\ref{eq:input}),
\begin{equation}
    I_n=\alpha U_0\int^\infty_{-\infty} dx
    \exp\left[-\frac{(x-z_0)^2}{4a^2}\right]v_n(x|z).
\end{equation}
After substituting Eq.~(\ref{eq:basis})
and completing square in the exponential argument, we have 
\begin{equation}
    I_n=\frac{\alpha U_0}{\sqrt{(2\pi)^{1/2}an!2^n}}
    \exp\left[-\frac{(z_0-z)^2}{8a^2}\right]
    \int^\infty_{-\infty} dx
    \exp\left[-\frac{(x-(z_0+z)/2)^2}{2a^2}\right]
    H_n\left(\frac{x-z}{\sqrt{2}a}\right).
\label{eqD2}
\end{equation}
Using the identity in Eq.~(\ref{eq:hx+y}),
denoting $(z_0+z)/2$ as $\bar z$,
and rewriting the integrand in terms of the basis functions, we obtain 
\begin{eqnarray}
    I_n &=&\frac{\alpha U_0}{\sqrt{(2\pi)^{1/2}an!2^n}}
    \exp\left[-\frac{(z_0-z)^2}{8a^2}\right]
    \int^\infty_{-\infty} dx \sqrt{(2\pi)^{1/2}a}v_0(x|\bar{z})
    \nonumber \\
    &\times & \sum_{k=0}^{n}\left(\begin{array}{c}
    k\\n\end{array}\right)\sqrt{(2\pi)^{1/2}ak!2^k}
    v_k(x|\bar{z})\left(\frac{z_0-z}{\sqrt{2}a}\right)^{n-k}.
\label{eqD4}
\end{eqnarray}
By the orthonormality of $v_n$,
only the $k=0$ term remains non-vanishing,
leading to the signal term in Eq.~(\ref{eq:stim_n}).

For the noise component,
\begin{equation}
    \eta_n(t)=\int^\infty_{-\infty} dx \eta(x,t)v_n(x|z).
\label{eqD6}
\end{equation}
It is straightforward to check that
\begin{eqnarray}
    \langle \eta_n(t) \rangle & = &
    \int^\infty_{-\infty} dx\langle\eta(x,t)\rangle v_n(x|z)=0,
\label{eqD7} \\
    \langle \eta_n(t)\eta_m(t) \rangle & = &
    \int^\infty_{-\infty} dx \int^\infty_{-\infty} dx'
    \langle \eta(x,t)\eta(x',t')\rangle v_m(x|z)v_n(x'|z)
    =\delta_{mn}\delta(t-t').
\label{eqD8}
\end{eqnarray}

\section*{Appendix F: The Dynamics of the 2D CANN}
\subsection*{F.1 The Eigenvalues and Eigenfunctions of $F(\vx,\vx'|\vz)$}

Define
\begin{equation}
    u_{m,n}(\vx|\vz)=u_m(x_1|z_1)u_n(x_2|z_2).
\end{equation}
We check that $u^{m,n}$ is the right eigenfunction of $F(\vx,\vx'|\vz)$
and calculate the corresponding eigenvalue.

We distinguish three different cases:
\begin{itemize}
\item $m=n=0$. Following the calculation in Eq.~(\ref{eqA3})
and noting the integrations with respect to $x'_1$ and $x'_2$
can be carried out separately, we have
\begin{eqnarray}
    \int^\infty_{-\infty} \int^\infty_{-\infty} d\vx'
    F(\vx,\vx'|\vz)u_{0,0}(\vx'|\vz) & = &
    \int^\infty_{-\infty} \int^\infty_{-\infty} dx'_1 dx'_2
    F(\vx,x'_1,x'_2|z_1,z_2)u_0(x'_1|z_1)u_0(x'_2|z_2) \nonumber \\
    & = & (1-\sqrt{1-k/k_c})u_{0,0}(\vx|\vz).
\end{eqnarray}
\item $m\neq 0, n=0$ (the case for $m=0$, $n\neq 0$ is the same).
An important observation
in the calculation of Eq.~(\ref{eqA2}) is that for $m\geq 1$,
the integral in the second term vanishes. Therefore
\begin{equation}
    \int^\infty_{-\infty} \int^\infty_{-\infty} d\vx'
    F(\vx,\vx'|\vz)u_{m,0}(\vx'|\vz)
    = \lambda_m u_{m,0}(\vx|\vz).
\end{equation}
\item
$m\neq 0, n\neq 0$.
\begin{equation}
    \int^\infty_{-\infty} \int^\infty_{-\infty} d\vx'
    F(\vx,\vx'|\vz)u_{m,n}(\vx'|\vz)
    = \lambda_m \lambda_n u_{m,n}(\vx|\vz).
\end{equation}
\end{itemize}

\subsection*{F.2 The network dynamics in the weak input limit}

In the limit of weak inputs, we assume that the shape of the bump 
is unchanged and only the position of the bump varies with
time, i.e., $U(\vx,t)\approx \tilde{U}(\vx|\vz(t))$.
We project the network dynamics in Eq.~(\ref{eq:dyn2D}) on a
single primary displacement mode, which is
the movement of the bump along the straight line
towards the stimulus location $\vz^0$.
Expressing the left-hand side of Eq.~(\ref{eq:dyn2D})
in terms of the basis functions leads to 
\begin{equation}
    \tau \frac{\partial}{\partial t}\tilde{U}(\vx|\vz,t)
    =\frac{\tau U_0 (2\pi)^{1/2}}{2}
    \left[u_{1,0}\frac{dz_1}{dt}+u_{0,1}\frac{dz_2}{dt}\right].
\end{equation}
On the right-hand side of Eq.~(\ref{eq:dyn2D}),
we note that the neuronal interaction and relaxation terms
cancel each other at the steady state,
leaving behind the external input.
Consider first the signal part,
projected onto the components $u_{1,0}$ and $u_{0,1}$: 
\begin{equation}
    R.H.S_{\rm signal}=\alpha U_0
    \int^\infty_{-\infty}\int^\infty_{-\infty} d\vx
    \exp\left[-{\frac{(\vx-\vz^0)^2}{4a^2}}\right]
    [u_{1,0}(\vx|\vz)+u_{0,1}(\vx|\vz)].
\end{equation}
Explicit integration yields
\begin{equation}
    R.H.S._{\rm signal}=-\alpha\frac{U_0 (2\pi)^{1/2}}{2}
    \exp\left[-{\frac {(\vz-\vz^0)^2} {8a^2}}\right]
    [(z_1-z_1^0)u_{1,0}(\vx|\vz)
    +(z_2-z_2^0)u_{0,1}(\vx|vz)].
\end{equation}
For the noise component,
\begin{equation}
    R.H.S._{\rm noise}=\sigma\eta_{1,0}(t)u_{1,0}(\vx|\vz)
    +\sigma\eta_{0,1}(t)u_{0,1}(\vx|\vz),
\end{equation}
where
\begin{eqnarray}
    \langle \eta_{1,0}(t) \rangle =
    \langle \eta_{0,1}(t) \rangle & = & 0, \\
    \langle \eta_{1,0}(t)\eta_{1,0}(t') \rangle =
    \langle \eta_{0,1}(t)\eta_{0,1}(t') \rangle
    & = & \delta(t-t').
\end{eqnarray}
Using the orthonormality of $u_{m,n}$ 
and expressing the result in vector form,
we arrive at Eq.~(\ref{eq:velocity2D}).

\newpage

\bibliographystyle{chi}
\bibliography{citation}

\end{document}